\documentclass{article}
\usepackage[utf8]{inputenc}
\usepackage{notoccite}
\usepackage{graphicx}
\usepackage{authblk}
\usepackage{appendix}

\title{INDIGO-DataCloud: a platform to facilitate seamless access to e-infrastructures}

\author[ ]{\large by the {{\bf INDIGO-DataCloud Collaboration} \\}} 

\author[1]{\small{\\ D. Salomoni}}
\author[2]{I. Campos}
\author[3]{L. Gaido}

\author[2]{J. Marco de Lucas}
\author[14]{P. Solagna}
\author[9]{J. Gomes}
\author[15]{L. Matyska}
\author[5]{P. Fuhrman}
\author[8]{M. Hardt}
\author[4]{G. Donvito}
\author[13]{L. Dutka}
\author[7]{M. Plociennik}
\author[11,21]{R. Barbera}

\author[6]{I. Blanquer}
\author[1]{A. Ceccanti}
\author[18]{E. Cetinic}
\author[9]{M. David}
\author[1]{C. Duma}
\author[2]{A. L\'opez-Garc\'{\i}a}
\author[6]{G. Molt\'o}
\author[2]{P. Orviz}
\author[15]{Z. Sustr}
\author[14]{M. Viljoen}

\author[2]{F. Aguilar}
\author[9]{L. Alves}
\author[4]{M. Antonacci}
\author[17]{L.A. Antonelli}
\author[3]{S. Bagnasco}
\author[16]{A.M.J.J. Bonvin}
\author[11]{R. Bruno}
\author[14]{Y. Chen}
\author[22]{F. Chiarello}
\author[17]{A. Costa}
\author[1]{S. Dal Pra}
\author[18]{D. Davidovic}
\author[22]{A. Dorigo}
\author[8]{B. Ertl}
\author[22]{F. Fanzago}
\author[11]{M. Fargetta}
\author[10]{S. Fiore}
\author[17]{S. Gallozzi}
\author[16]{Z. Kurkcuoglu}
\author[2]{L. Lloret}
\author[9]{J. Martins}
\author[10]{A. Nuzzo}
\author[10]{P. Nassisi}
\author[10]{C. Palazzo}
\author[9]{J. Pina} 
\author[17]{E. Sciacca}
\author[22]{M. Segatta}
\author[22]{M. Sgaravatto}
\author[12]{D. Spiga}
\author[1]{S. Taneja}
\author[19]{M. Tangaro}
\author[7]{M. Urbaniak}
\author[3]{S. Vallero}
\author[22]{M. Verlato}
\author[8]{B. Wegh}
\author[3]{V. Zaccolo}
\author[19,20]{F. Zambelli}
\author[22]{L. Zangrando}
\author[1]{S. Zani}
\author[7]{T. Zok}

{\small {

\affil[1]{\small{INFN - CNAF, Bologna, Italy}}
\affil[2]{IFCA, Consejo Superior de Investigaciones Cientificas-CSIC, Santander (Spain)}
\affil[3]{INFN - Torino, Torino, Italy}
\affil[4]{INFN - Bari, Bari, Italy}
\affil[5]{Deutsches Elektronen Synchrotron (DESY), Germany}
\affil[6]{Institute of Instrumentation for Molecular Imaging - Universitat Politècnica de València, Spain}
\affil[7]{PSNC IBCh PAS, Poland}
\affil[8]{Karlsruhe Institute of Technology (KIT), Germany}
\affil[9]{Laboratory of Instrumentation and Experimental Particle Physics (LIP), Portugal}
\affil[10]{Fondazione Centro Euro-Mediterraneo sui Cambiamenti Climatici , Lecce, Italy}
\affil[11]{INFN - Catania, Catania, Italy}
\affil[12]{INFN Division of Perugia, Italy}
\affil[13]{Cyfronet AGH, Krakow, Poland}
\affil[14]{EGI Foundation, Amsterdam (Netherlands)}
\affil[15]{CESNET, Prague, Czech Republic}
\affil[16]{University of Utrecht, The Netherlands}
\affil[17]{Istituto Nazionale di Astrofisica, Italy}
\affil[18]{Ruder Boskovic Institute, Zagreb (Kroatia)}
\affil[19]{Istituto di Biomembrane, Bioenergetica e Biotecnologie Molecolari, Consiglio Nazionale delle Ricerche, Bari, Italy}
\affil[20]{University of Milano, Dept. of Biosciences, Milan, Italy}
\affil[21]{Department of Physics and Astronomy, Univ. of Catania, Italy }
\affil[22]{INFN - Padova, Padova, Italy }
}}

\begin{document}

\maketitle

\newpage

\begin{abstract}
This paper describes the achievements of the H2020 project INDIGO-DataCloud. The project has provided  e-infrastructures  with  tools, applications and cloud framework enhancements to manage the demanding requirements of scientific   communities,   either   locally   or through enhanced  interfaces. The middleware developed allows to federate hybrid resources, to easily write, port and run scientific applications to the cloud. Our developments are freely downloadable as open source components, and are already being integrated into many scientific applications.
\end{abstract}

\newpage

\tableofcontents

\newpage

\section{Introduction}

INDIGO-DataCloud was an European project starting in April 2015, with the
purpose of developing a modular architecture and software components to improve
how scientific work is supported at the edge of computing services development.
Its main goal has been to deliver a Cloud platform addressing the specific
needs of scientists in a wide spectrum of disciplines, engaging public
institutions and private companies. It aimed at being as inclusive as possible,
developing open source software exploiting existing solutions, adopting and
enhancing state of the art technologies, connecting with other initiatives and
with leading commercial providers.

Since its inception, the project roadmap has been user community driven. Its
main focus was on closing the existing technology gaps that hindered an
optimal exploitation of Cloud technologies by scientific users. In order to do
so, user requirements from several multidisciplinary scientific communities
were collected, and systematized into specific technical requirements. This
process was carried out across the entire lifetime of the project, which
allowed the update of existing requirements as well as the insertion of new
ones, thus driving the project architecture definition and the technological
developments.

The project also made focus on delivering production-quality software, thus it
defined procedures and quality metrics, which were followed by, and
automatically checked for, all the INDIGO components. A comprehensive process
to package and issue the INDIGO software was also defined. As an outcome of
this, INDIGO delivered two main software releases (the first in August 2016,
the second in April 2017), each followed by several minor updates. The latest
release consists of about 40 open modular components, 50 Docker
containers, 170 software packages, all supporting up-to-date open operating
systems. This result was accomplished reusing and extending open source
software and ---whenever applicable--- contributing code to upstream projects.

The paper is structured as follows.
Section~\ref{sec:user-req} contains a description on how the user requirements
were collected and consolidated. From there, the INDIGO architecture is further
elaborated from the lower Infrastructure as a Service layer
(Section~\ref{sec:iaas}) moving towards the Platform layer
(Section~\ref{sec:paas}) in order to arrive to the user interfaces
(Section~\ref{sec:users}). The overall software development process is
described in Section~\ref{sec:release}. Section~\ref{sec:examples} contains a
summary of some usage patterns on how to leverage the INDIGO solutions to
develop, deploy and support applications in a Cloud framework. The conclusions
are laid out in Section~\ref{sec:conclusions}. The list of upstream contributed
software can be found in the Appendix.

\subsection{Context and state of the art}

From the collection of user community requests, and its consolidation into technical requirements (see section \ref{sec:user-req}), we identified a number of technology gaps that today hinder an optimal scientific exploitation of heterogeneous e-infrastructures.

In this Section we will elaborate more on the general strategy to address those requirements, linking our developments with the previous and existing works. The specific enhancements and developments will be further elaborated in the corresponding sections.

Lack of proper federated identity support across several e-Infrastructures is a key issue for the researchers perspective. The provision of an effective distributed authentication and authorization in heterogeneous platforms is fundamental to support access to distributed infrastructures. Several efforts have been made in this context
\cite{LopezGarcia2013b,Chadwick2014,lee2014design,pustchi2015authorization,lee2014keystone}
but they were focused on specific infrastructures and services. However,
although some of these approaches have been used in production in specific
e-Infrastructures \cite{FernandezdelCastillo2015} they are difficult 
to implement in a broader environment. 

In parallel to the development of INDIGO-Datacloud, the
{\sl Authentication and Authorisation for Research and Collaboration} project (AARC)
defined the AARC Blueprint Architecture \cite{aarc}. This document describes a 
set of interoperable architecture building blocks for designing and implementing access management
solutions for international research collaborations. Following the AARC recommendations we have
developed several key components related with identity and access management,
providing a framework compliant with the proposed blueprint architecture, as
will be described in Section~\ref{sec:iam}.

Facilitating the transparent execution of user applications across different
computing infrastructures is also a key issue \cite{Oesterle2015}. 
Advanced users have nowadays at their disposal tools to implement 
applications in Clouds provisioned in {\sl Infrastructure as a Service} (IaaS) mode. 
Examples of such solutions are virtual appliances and contextualization \cite{Campos2013} or 
container technologies \cite{boettiger2015introduction}). 

The situation for non-Cloud resources in scientific facilities is completely different.
Here we are referring for instance to local clusters, Grid infrastructures and HPC systems. 
Such infrastructures are tipically shared among many users with different requirements, therefore 
it is managerially and technically impossible offering tailored environments to all of them.
As a consequence scientific users often need to follow a troublesome process to package and 
execute their applications. 

To address this problem we have applied the technology of Docker containers \cite{DOCKER} to facilitate applications execution in multiuser environments. As a result we have provided a flexible user-level solution to provide autonomy to users in shared computing facilities \cite{udocker-paper}. Section \ref{sec:containers} contains a thorough discussion on the strategy and outcomes.

Adoption of true Platform as a Service (PaaS) Cloud solutions is a common problem for
scientific communities. The roots of this problem are on the one hand the non-interoperability of the interfaces \cite{Zhang2013,Lorido-Botran2014}, and second, the lack of true orchestration
mechanisms across federated heterogeneous infrastructures. Both barriers made it difficult 
for users to adopt Cloud hybrid solutions.

In Section~\ref{sec:paas} we describe our approach, and how we have tackled this problem by 
leveraging the OCCI \cite{Nyren2010,Metsch2010,Metsch2011} and TOSCA \cite{TOSCA} open
standards. In this regard we have not only supported those standards at the
corresponding architectural levels \cite{Teckelmann2011,LopezGarcia2016b}, but
also we made important contributions to both the standards specifications and
implementations. INDIGO has contributed to the networking parts of the OCCI
standard, as well as to the improvement of the TOSCA support in the upstream OpenStack
components: the Heat Translator and TOSCA parser \cite{LopezGarcia2017b}. Our solution makes the execution of dynamic workflows \cite{Hardt2012,Fakhfakh2014,Stockton2017} possible, in more consistent way across hybrid Clouds \cite{PLOCIENNIK2016722}.

In this interoperability context, hybrid Cloud deployments, although possible
\cite{moreno2011multicloud,Katsaros,Lorido-Botran2014}, were complicated from a practical 
point of view and therefore user adoption has been hindered. 
By adopting INDIGO solutions users can now express their requirements and deploy them as applications over those hybrid infrastructures \cite{LopezGarcia2017}. 

Linked with the previous statements another outstanding gap was the lack of
advanced scheduling features in Cloud environments
\cite{Singh2016}. Common cloud usage scenarios, being industry driven, do not
take into account the unique requirements of scientific applications
\cite{LopezGarcia2017c}, leading to an inefficient utilization of the resources
or to non optimal user experience. 

Developments in this area can be found in the literature 
\cite{Chauhan2017,Somasundaram2014,Sotomayor2006,Manvi2014}, where it becomes evident that 
there are many challenges to be addressed. Within INDIGO we focused
(see Section~\ref{sec:sched}) in the efficient sharing of resources between users
following fair-share approaches (limiting the amount of resources that can be
consumed by a user group), proper quota partitioning across different computing
frameworks (like HPC and Cloud resources) or new Cloud computing execution
models (like preemptible instances) as these are aspects hat affect both users
and resource providers.

Regarding storage support, INDIGO has performed substantial contributions to
storage-related entities and standardization bodies, such as the Research Data
Alliance (RDA), where INDIGO has been highly involved the Quality of Service,
Data Life cycle and Data Management Plans working group (now renamed to Storage
Service Definitions). Moreover, INDIGO has also contributed this work to the
SNIA CDMI standard, providing several extensions that have been included in
SNIA reference implementations and documents.


\section{Analysis of requirements coming from research communities}
\label{sec:user-req}

In order to guide our developments we performed an analysis of a number of use cases originating in several flagship research communities. In particular coming from the areas of High Energy Physics, Environmental modelling, Bioinformatics, Astrophysics and Social sciences. See Table \ref{tab:req} for the full list.


\begin{table}
  \centering
  \begin{tabular}{c|p{9cm}}
    {\bf Research community} & {\bf Application/Use Case} \\ \hline \hline
 \\   LIFEWATCH (Biodiversity) & {\bf Monitoring and Modelling Algae Bloom in a Water Reservoir:} Support of hydrodynamic and water quality modelling including data input-output management and visualization. \\ \hline
 \\   INSTRUCT (Bioinformatics) &  {\bf Molecular dynamics simulations:} Support of Molecular dynamics simulations of macromolecules that need specific hardware (GP/GPUs) using a pipeline of software that combines protocols that automate the step for setup and execution of these simulations.  \\ \hline
 \\   CTA (Astronomical Data) & {\bf Astronomical Data Archives:} Data analysis and management using different tools such as data discovery, comparison, cross matching, data mining and also workflows. The use case could be described as follows: data production, data reduction, data quality, data handling and workflows, data publication and data link to articles   \\ \hline
 \\   Climate Modelling &  {\bf Intermodel comparison} of data analysis for different climate models using the ENES platform (European Network for Earth System modelling) \\ \hline
 \\   EuroBioImaging (Bioinformatics) & {\bf Medical Imaging Biobanks:} The virtual Biobank integrates medical images from different sources and formats. This case study includes all the steps needed to manage the images, like analysis, storage, processing (pre, post). Privacy is a constraint to take into account for user management \\ \hline
 \\   ELIXIR (Bioinformatics) & {\bf Galaxy as a Cloud service:} Deployment of Galaxy instance that should support all the software/steps needed by the pipeline over, for example, a virtual cluster or cloud instances. \\ \hline
 \\   DARIAH (Social Sciences) & Transparent access to data catalogues and on-demand data management features. \\ \hline
 \\   Mastercode (HEP Pheno) & {\bf Complex combination of codes including legacy parts} to perform combined analysis of data coming from particle detectors, astrophysics experiments, and dark matter detectors. Installation of these codes is in general very complex in multi-user farms. Providing a container based solution would simplify installation across infrastructures. \\ \hline
 \\   Lattice QCD  (HEP) & Lattice simulations run on large HPC facilities using low latency interconnects, producing large amounts of output. Accessing such facilities in Cloud mode would require implementing {\bf MPI parallel processing capabilities}. \\ \hline
  \end{tabular}
  \caption{Research Communities and use cases analyzed to extract general requirements}
  \label{tab:req}
\end{table}

The deployment of customized computing back-ends, such as batch queues, including automatic elasticity is among the features more demanded by researchers. The automation of the deployment of user-specific software in VMs or containers is also on the top of their wish list. Such automation is a {\sl must} when it is about simplifying the executing applications in heteregeneous infrastructures. For similar reasons, highly specialized applications require also support to hardware accelerators and specialized hardware such as Infiniband, multicore systems, or GP/GPUs.

Often user communities are asking about terminal access to resources, workflow management and data handling, in a way that such access is linked to a common Authorization and Authentication Infrastructure. 

In order to generalize the requirements, we have extracted two generic usage scenarios, which can support a wide range of applications in these areas. The first generic use case is computing oriented, while the second is data analysis oriented. For full details regarding user communities description and detailed usage patterns we refer to the users requirements deliverable of the project available publicly in \cite{D21}.

\subsection{Computing Portal Service}

The first generic user scenario is a computing portal service. In such scenario, computing applications are stored by the application developers in repositories as downloadable images (in the form of VMs or containers). Such images can be accessed by users via a portal, and require a back-end for execution; in the most common situation this is typically a batch queue. Support for parallel processing using containers is a requeriment that comes up as well from the users.

\begin{figure}
  \centering
  \includegraphics[width=11cm]{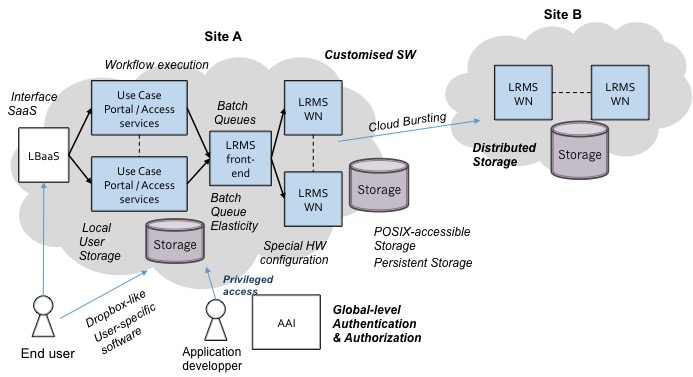}
  \caption{User Community Computing Portal Service}
  \label{fig:01}
\end{figure}

The application consists of two main parts: the portal / Scientific Gateway and the processing working nodes. The number of nodes available for computing should increase (scale out) and decrease (scale in), according to the workload. The system should also be able to do Cloud-bursting to external infrastructures when the workload demands it. Furthermore, users should be able to access and reference data, and also to provide their local data for the runs. A solution along these lines is shown in Figure \ref{fig:01}.

A solution along these lines has been requested in the user scenarios coming from ELIXIR, WeNMR, INSTRUCT, DARIAH, Climate Change and LIFEWATCH.

\subsection{Data Analysis Service}

A second generic use case is described by scientific communities that have a coordinated set of data repositories and software services to access, process and inspect them. 

\begin{figure}
  \centering
  \includegraphics[width=11cm]{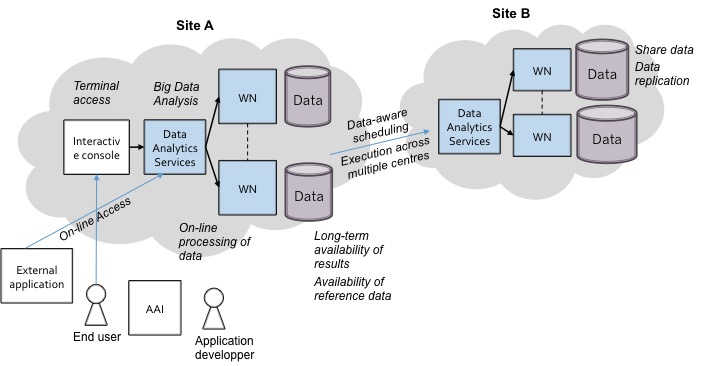}
  \caption{Data Analysis Service}
  \label{fig:02}
\end{figure}

Processing is typically interactive, requiring access to a console deployed on the data premises. The application consists of a console / Scientific Gateway that interacts with the data. In Figure \ref{fig:02} we show a schematic view of such a use case. Examples of such include {\sl R}, {\sl Python} or {\sl Ophidia}. It can be a complementary scenario to the previous one, and it could also expose programmatic services.

The communities related to INSTRUCT, CTA, Climate Change, LIFEWATCH and Lattice QCD have requested related features.

\section{Developing for the Infrastructure as a Service (IaaS) Layer}
\label{sec:iaas}

INDIGO-DataCloud  has  provided  e-infrastructures  with  tools, applications and cloud framework enhancements to manage the demanding requirements of modern scientific   communities,   either   locally   or through enhanced  interfaces enabling those infrastructures  to  become  part  of  a  federation  or to connect  to  federated  platform  layers (PaaS). 

In this section we will describe the highlights of this development work, which was needed to properly interface with the resource centers. This work has focussed on virtualizing local compute, storage and networking resources (IaaS) and on providing those resources in a standardized, reliable and performing way to remote customers or to higher level federated services, building virtualized site independent platforms.

The IaaS resources are provided by large resource centers, typically engaged in well-established European e-infrastructures. The e-infrastructure management bodies, or the resource centers themselves will select the components they operate, and INDIGO will have limited influence on that process. 

Therefore, INDIGO has concentrated on a selection of the most prominent existing components and has further developed the appropriate interfaces to high-level services based on standards. We have also developed new components where we felt a full functionality was completely missing. 

The contribution of INDIGO to enhance the flexibility to access resources in Cloud and HPC infrastructures will be of paramount  importance to  enable  transparent  execution  of  applications  across  systems promoting the development of the future European Open Science Cloud (EOSC) ecosystem \cite{EOSC}.

As we describe below new components  are  provided,  or already  existing  components  are improved  in  the areas   of   computing,   storage,   networking and Authorization and Authentication Infrastructure (AAI).   For   almost   all   components, we succeeded in committing modifications to the corresponding upstream software providers and by that, significantly contributed to  the  sustainability  model  of  the  software.

\subsection{Supporting Linux containers}
\label{sec:containers}

It is unquestionable that Docker is the most widely adopted Linux container technology. Therefore, in order to facilitate application delivery across multiple computational platforms, INDIGO  has  provided  support  for  container  execution, both  interactively  and  through  batch systems,  in cloud and conventional clusters. This was achieved by developing new tools and extending existing ones.
 
The key middleware developed for this purpose is {\bf udocker} \cite{udocker-paper}\footnote{https://github.com/indigo-dc/udocker}. The udocker novelty consists in enabling to pull and execute Docker containers \cite{DOCKER} without using or requiring the installation of the Docker software. By using udocker it is possible to encapsulate applications in Docker containers and execute them in batch or interactive systems where Docker is unavailable. It provides several different execution engines based on PRoot\cite{PROOT}, runC\cite{RUNC} and Fakechroot\cite{FAKECHROOT}. None of the udocker engines requires root privileges for installation or execution being therefore adequate for deployment and use by end-users without system administrator intervention. In addition, the PRoot and Fakechroot engines execute containers via pathname translation and therefore do not require the use of Linux namespaces. 

Since udocker never requires privileges and executes as unprivileged user many of the security concerns associate with the Docker software are avoided. Udocker also supports GPGPU and MPI applications, making it adequate to execute containers in batch systems and HPC clusters. The udocker software suite is meant to be easily deployed by end-users. It only requires the download and execution of a Python script to quickly setup udocker within the user home directory. Udocker empowers end-users to execute Docker containers regardless of the Linux host environment.

Since its first release in June 2016 udocker expanded quickly in the open source community. It has been adopted by a number software projects as a drop-in replacement for Docker. Among them openmole, bioconda, common-work language (cwl)  or SCAR - Serverless Container-aware ARchitectures \cite{Perez2018scc}.

As an example, udocker  is  being  used  with  great  success  to  execute  code  produced  by  the  MasterCode collaboration \cite{MASTERCODE}.   The   MasterCode   collaboration   is   concerned   with   the   investigation of Supersymmetric  models  that  go  beyond  the  current  status  of  the  Standard  Model  of  particle physics.  It  involves  teams  from  CERN,  DESY,  Fermilab,  SLAC,  CSIC,  INFN,  NIKHEF, Imperial College London,   King's   College   London,   the   Universities   of   Amsterdam, Antwerpen, Bristol, Minnesota and ETH Zurich. Examples and documentation can be found at https://github.com/indigo-dc/udocker.

INDIGO has also developed {\bf bdocker}\footnote{https://github.com/indigo-dc/bdocker}, which provides a front-end to execute the Docker software in batch systems under restrictions and limits configurable by the system administrator (resource consumption, access to host directories, list of container images, etc). It has been implemented for the SoGE (Son of Grid Engine) batch system but can be extended to other batch systems. This integrates the benefits of application delivery provided by Docker with the scheduling policies of the batch system. While udocker can be deployed directly by the end-user, bdocker is installed and configured by the system administrator to provide the users with the ability to run limited execution environments provided by Docker. Finally, ONEDock\footnote{https://github.com/indigo-dc/onedock} was developed to introduce support to the execution of containers as if they were Virtual Machines in OpenNebula-based on-premises Clouds, by supporting Docker as a hypervisor in this Cloud Management Platform. With this approach, applications and their execution environment packaged as Docker images can be instantiated on-demand through OpenNebula and provide SSH-based access for multiple users, with the advantage for the administrators of reduced overhead (such as low memory footprint) when compared to Virtual Machines.

\subsection{Development of advanced Scheduling Technologies}
\label{sec:sched}

The end goal of this activity in INDIGO is improving the performance of the cloud management platforms by designing and implementing novel scheduling mechanisms and policies at the IaaS level. Enabling advanced scheduling policies to optimize the usage of the data center will clearly improve the response to the users. 

To this end cloud schedulers need to include support for postponing low priority workloads (by killing, preempting or stopping running containers or VMs) in order to allocate higher priority requests. 


\subsubsection{Preemptible Instances}

INDIGO pushed the  state  of  the  art  in  scheduling  technologies  by  implementing  preemptible instances  on  top  of  the  OpenStack\cite{OPENSTACK}  cloud  management  framework, {\bf opie}\footnote{https://github.com/indigo-dc/opie}. Openstack preemptible instances is the materialisation of the preemptible instances extension, serving as a reference implementation.

Preemptible instances differ from regular ones in that they are subject to be terminated by a incoming request for provision of a normal instance. If bidding is in place, this special type of instance could also be stopped by a higher priority preemptible instance (higher bid). Not all the applications are suitable for preemptible execution, only fault-tolerant ones can withstand this type of execution. On the other side they are highly affordable VMs that allow providers to optimize the usage of their available computing resources (i.e. backfilling).

The opie package provides a set of pluggable extensions for OpenStack Compute (nova) making possible to execute preemptible instances using a modified filter scheduler. This solution has gained great interest from the scientific community and commercial partners, and is under discussions to  be  introduced  in  the  upstream  OpenStack  scheduler.  

\subsubsection{Implementing advanced scheduling in OpenStack and OpenNebula}

In IaaS private clouds the computing and storage resources are statically partitioned among projects. A user typically is member of one project, and each project has its own fixed quota of resources defined by the cloud administrator. A user request is rejected if the project quota has been already reached, even if unused resources allocated to other projects would be available. 

This rigid resource allocation model strongly limits the global efficiency of the data centres, which aim to fully utilize their resources for optimizing costs. In the traditional computing clusters the utilization efficiency is maximized through the use of a batch system with sophisticated scheduling algorithms plugged in. However this feature is missing in the most popular cloud middlewares. In the course of INDIGO we have developed support for advanced scheduling policies such as intelligent job allocation based on fair-share algorithms for both OpenStack and OpenNebula cloud frameworks.

\begin{itemize}
\item  {\bf Synergy}\footnote{https://github.com/indigo-dc/synergy-service} is an advanced service interoperable with the OpenStack components, which implements a new resource provisioning model based on pluggable scheduling algorithms. It allows to maximize the resource usage, at the same time guaranteeing a fair distribution of resources among users and groups. 

The service also provides a persistent queuing mechanism for handling those user requests exceeding the current overall resource capacity. These requests are processed according to a priority defined by the scheduling algorithm, when the required resources become available.

\item The scheduling capabilities of OpenNebula have been enhanced with the development of {\bf one-FaSS} (FairShare Scheduler for OpenNebula)\footnote{https://github.com/indigo-dc/one-fass}. In OpenNebula the scheduler is first-in-first-out (FIFO). One-FaSS grants fair access to dynamic resources priorizing tasks assigned according to an initial weight and to the historical resource usage.

\end{itemize}

The project has also developed tools to facilitate the management of hybrid data centers, this is, where both batch system based and cloud based services are 
provided. Physical computing resources can play both roles in a mutual exclusive way. The {\bf Partition Director}\footnote{https://github.com/indigo-dc/dynpart} takes care of commuting the role of one or more physical machines from {\sl Worker Node} (member of the batch system cluster) to {\sl Compute Node} (member of a cloud instance) and vice versa.

The current release only works with the IBM/Platform LSF Batch system (version 7.0x or higher) and Openstack Cloud manager instances (Kilo or newer).
The main functionalities are switch role of selected physical machines from the LSF cluster to the Openstack one and viceversa, and manage intermediate transition status to ensure consistency.

\subsection{Development of Authorization and Authentication Infrastructures}
\label{sec:iam}

INDIGO  has  provided  the  necessary components to offer  a commonly  agreed  Authentication  and  Authorization  Infrastructure  (AAI). The INDIGO {\bf IAM}\footnote{https://github.com/indigo-dc/iam} (Identity and Access Management service) provides user identity and policy information to services so that consistent authorization decisions can be enforced across distributed services.

IAM has a big impact on the end-user experience. It provides a layer where identities, enrollment, group membership and other attributes and authorization policies on distributed resources can be managed in a homogeneous way, supporting the federated authentication mechanisms supported by the INDIGO AAI. 

The INDIGO  AAI solution pioneers the usage of OpenID  Connect (OIDC) on  the  SP-IdP  proxy. INDIGO has made  contributions  to  the  upstream  components  whenever  needed  to  enable OpenID  Connect (namely  in  OpenStack  Keystone  and  Apache  Libcloud). 

\begin{figure}
  \centering
  \includegraphics[width=10cm]{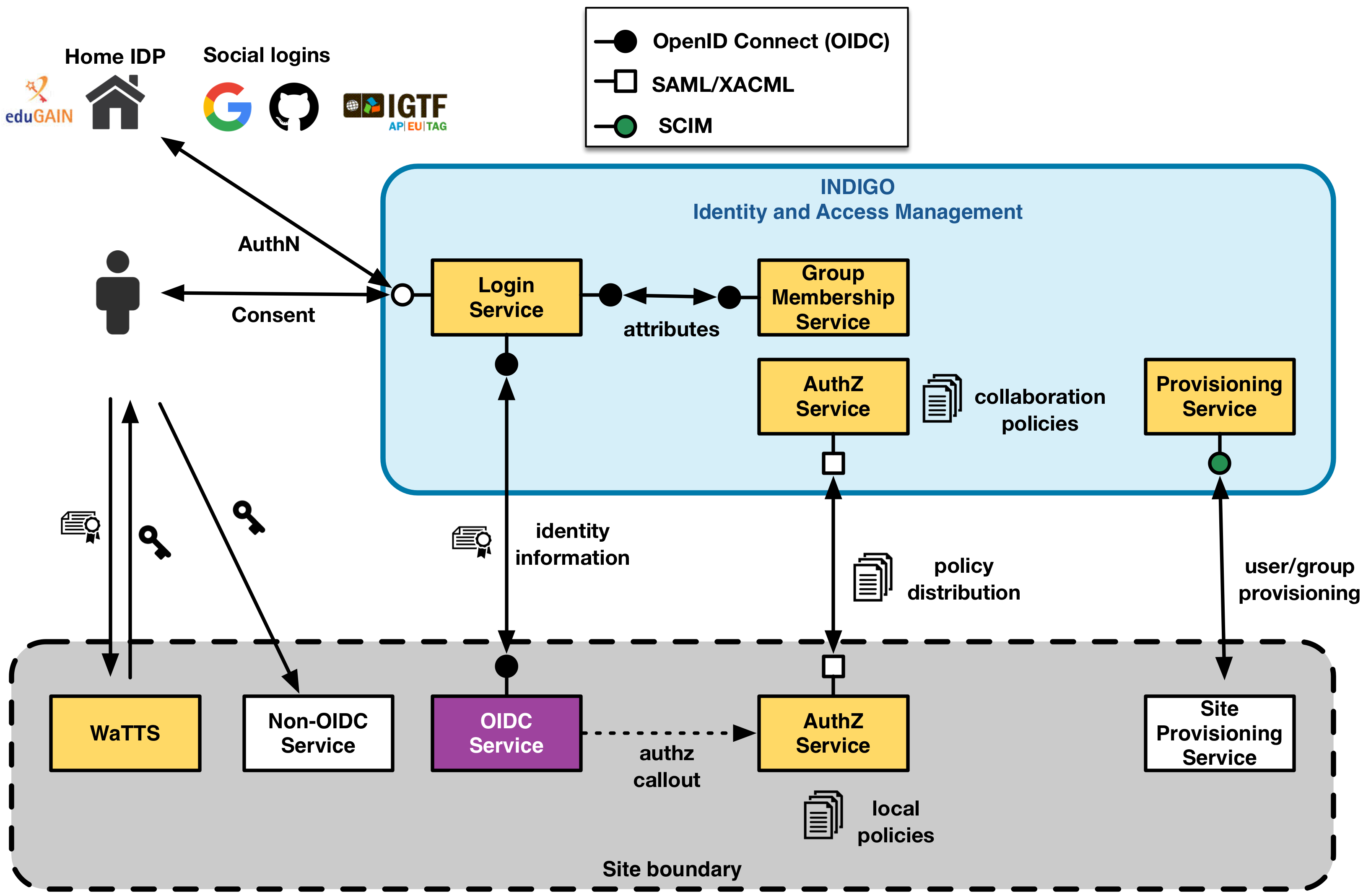}
  \caption{Architecture of the INDIGO Identity and Access management Service}
  \label{fig:0}
\end{figure}

INDIGO DataCloud provides a flexible Authentication and Authorization Infrastructure (AAI) whose main components are depicted in Figure \ref{fig:0}.

In order to do authentication and authorization in a consistent way, services
rely on the information provided by the central IAM service.

The Login Service component implements brokered authentication: users can
authenticate with any of the supported mechanisms (SAML, OpenID Connect, X.509
certificates, local username/password). Identity and authorization information
is then exposed to services via standard OAuth/OpenID Connect protocols. This
approach simplifies integration with off-the-shelf components and do not
overload all services in the infrastructure with the complexity of handling
multiple credential types. This approach has a big impact on end-user
experience, as it allows users to centralize the management of their
credentials and provide a consistent login experience to all services in the
system.

The Group Membership Service component provides the tools to manage
registration and enrollment flows for the collaboration, organize users into
groups and manage user account life cycle.

The Authorization Service component, based on the Argus Authorization 
Service \cite{ARGUSAS}, 
provides the ability to define fine-grained policies for the collaboration
leveraging the flexibility of a XACML policy engine \cite{XACML}. This component also provides a policy composition and distribution mechanism that is used to ensure consistent authorization across the distributed infrastructure.

The Provisioning Service component exposes an SCIM API \cite{SCIM} to provision
information about the collaboration users to relying services. This mechanism
is useful, for instance, to manage the lifecycle of resources at a site (e.g.,
local UNIX accounts) depending on the lifecycle of IAM user account information. 
As an example, accounts could be provisioned automatically across the
infrastructure and configured with user SSH public keys as registered in the
central IAM at user registration time, and disabled when the user membership at
the IAM expires or is suspended due to a security incident. 

Finally, to integrate services that do not speak OpenID Connect natively, IAM
integrates with WaTTS, the INDIGO Token Translation Service\footnote{https://github.com/indigo-dc/tts}. WaTTS can translate 
identity and authorization information about a user provided as OpenID Connect
tokens to various credential types. This allows the provision of services that
do not normally support federated identities to federated users.

IAM is used internally to the PaaS layer in order to deal with the authorization of each user to the services, but also in handling group membership and role management for each user. Users may present themselves with any of the three supported token types (X.509, OpenID Connect, SAML) and are able to access resources that support any of them.

\subsection{Virtual Networks}

The INDIGO PaaS layer has been developed to exploit a wide range of cloud management frameworks (e.g. OpenStack, OpenNebula, Google Compute Engine, Microsoft Azure, Amazon EC2) and combine resources provided by these frameworks to enable the deployment of complex distributed applications. Each cloud management framework may exhibit a different native API and often these APIs can be configured in different ways. This heterogeneity constitutes a challenge when transparent instantiation of cloud resources across multiple frameworks is required. 
The INDIGO PaaS layer supports both common native APIs as well as the {\bf Open Cloud Computing Interface}\footnote{http://occi-wg.org/} ({\tt OCCI}). The OCCI specification is a standard from the {\bf Open Grid Forum}\footnote{https://www.ogf.org} that provides a flexible and extensible API to access and manage cloud resources. Although OCCI provides a convenient uniform API, its support to manage the network environment is limited in what concerns the setup of public/private network accessibility. Depending on the actual cloud management framework being used the target cloud may need manual network configuration prior to the use of OCCI. To address these problems INDIGO has defined and implemented an OCCI network extension that allows the network environment to be properly setup via the OCCI API regardless of the underlying cloud management framework.

For OpenNebula\cite{OPENNEBULA} sites the solution consists in a {\bf Network Orchestrator Wrapper} (NOW) \footnote{https://github.com/indigo-dc/now} and a corresponding backend in the {\bf rOCCI-server}\footnote{https://github.com/the-rocci-project/rOCCI-server}. NOW enforces site-wide policy and network configuration by making sure that only LANs designated by site administrators are made available to users, and that users cannot reuse LANs assigned to others while they remain reserved. NOW has been released with INDIGO, and the backend has been provided as a contribution to upstream rOCCI-server distribution.

For OpenStack, the OpenStack OCCI Interface (OOI) has been extended with support for advanced networking functions provided by OpenStack’s Neutron component such as router, network and subnet setup. The contribution was accepted upstream and is distributed with the OOI implementation.

The networking features of the OCCI gateway for the Amazon’s EC2 API were adjusted making sure that the model of setting up and using local virtual networks is in accordance with the model used in the other cloud management frameworks.

In addition a Virtual Router was implemented allowing networks to span across cloud  sites,  potentially geographically  distant,  so  that  a custom  networking  environment  can  be  setup  even if the resources are allocated in different cloud sites. The virtual router is a virtual machine that can be started via OCCI and makes use of {\bf OpenVPN}\footnote{https://openvpn.net} to implement network tunnels. The Virtual Routers can be instantiated by the PaaS layer to orchestrate the interconnection of virtual machines across cloud providers.


\section{Architecture of the Platform as a Service}
\label{sec:paas}

Generally speaking, a Platform as a Service (PaaS) is a software suite, which is able to receive programmatic resource requests from end users, and execute these requests provisioning the resources on some e-infrastructures. We can see already many examples in the industrial sector, in which open source PaaS solutions (eg. OpenShift\cite{OPENSHIFT} or Cloud Foundry\cite{CLOUDFOUNDRY} are being deployed to support the work of companies in different sectors.

The case of supporting scientific users is more complex in general than supporting commercial activities, because of the heterogeneous nature of the infrastructures at the IaaS level (i.e. the resource centers) and of the inherent complexity of the scientific work requirements. The key point is to find the right agreement to unify interfaces between the PaaS and IaaS levels.

The {\bf Infrastructure Manager}\cite{IM}(IM) has been used to address the IaaS level. The IM is able to deploy complex and customized virtual infrastructures on IaaS Cloud deployment(such as AWS, OpenStack, etc.). It automates the deployment, configuration, software installation, monitoring and update of the virtual infrastructure on multiple Cloud back-ends. 

In the framework of INDIGO the IM has extended its capabilities. In particular, in the PaaS it is used by the Orchestrator (see below) in order to provision and configure the virtual infrastructure required to support the scientific applications involved in the project\footnote{https://github.com/indigo-dc/im}.

In order to better adapt to the wide range of use cases provided by the users communities we decided to take a different approach from many of the more used PaaS: our solution is based on the concept of orchestrating complex cluster of service and on the possibility to automatize the actions needed to implement the use cases. This approach was really successful as it gave the possibility to implement also legacy applications and did not depended on the language in which the application is built. 

\begin{figure}
  \centering
  \includegraphics[width=\textwidth]{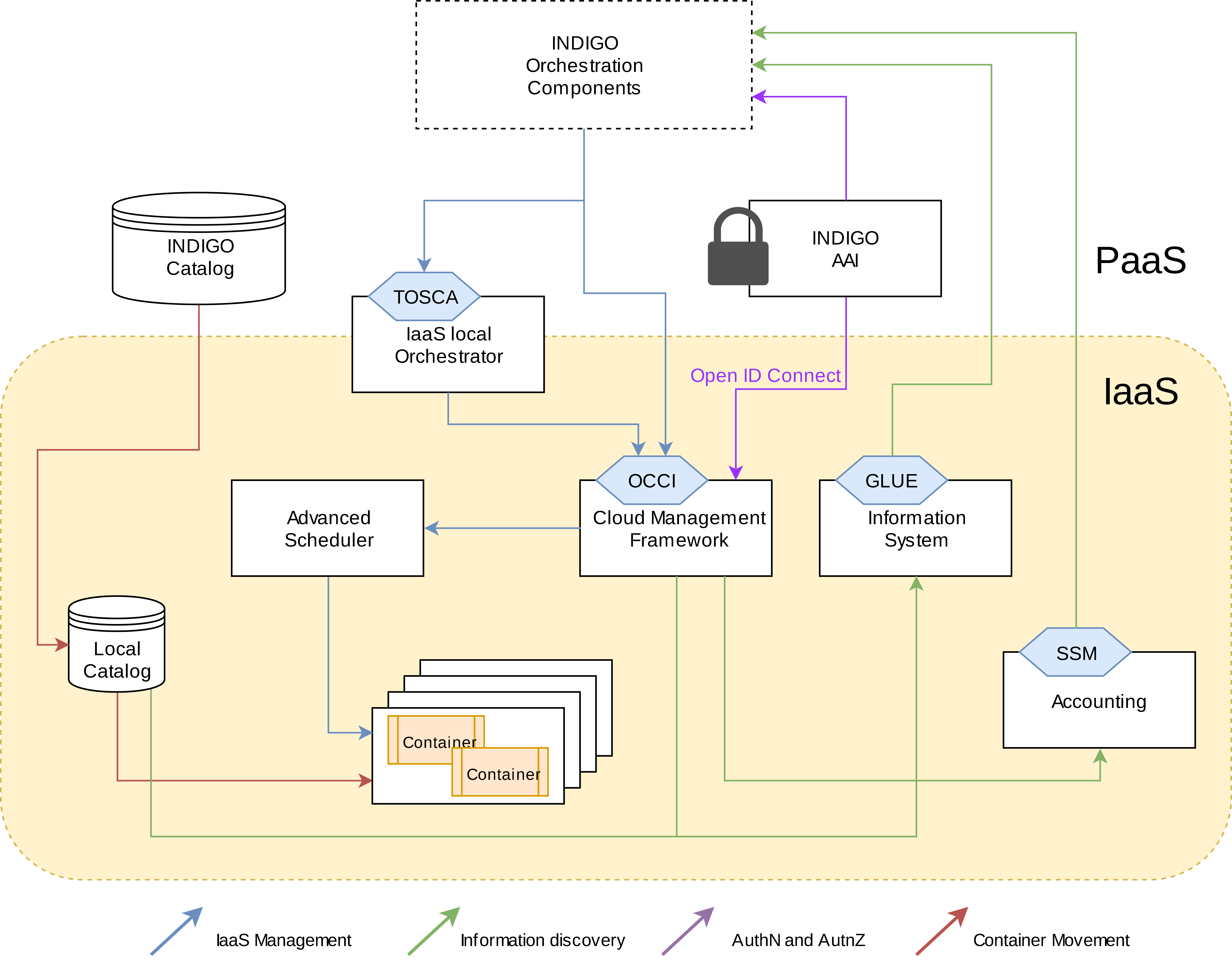}
  \caption{Interaction between the IaaS and PaaS layers}
  \label{fig:2}
\end{figure}

In Figure \ref{fig:2} we show the general interaction between the IaaS and PaaS layers. The Orchestrator provides the entry point to the PaaS layer with its ability to decide the most appropriate site on which to deploy a certain application architecture described in TOSCA Templates. INDIGO-DataCloud fosters local-site orchestration and, therefore, depending on the underlying Cloud Management Framework of the Cloud site, the TOSCA Template is translated the specific orchestration component of OpenStack (Heat) or it is delegated on the Infrastructure Manager (IM) to execute on OpenNebula-based Cloud sites. Since both Virtual Machines and containers can be provisioned on the underlying Cloud site, Virtual Machine Images available in each are registered in the Information System and container images are pre-staged to the Cloud sites to reduce deployment times.

INDIGO has provided a working PaaS Layer orchestrating heterogeneous computing and storage resources. Using the PaaS Orchestrator together with the IM and TOSCA Templates, the end users are able to exploit computational resources without knowledge about the IaaS details. In the following we describe the main technologies employed to build the PaaS.  

\subsection{PaaS layer and microservices architecture}

The Paas layer should be able to hide complexity and federate resources for both Computing and Storage. For that we have applied the current technologies based on lightweight containers and related virtualization developments using microservices. 

Kubernetes\cite{KUBERNETES}, an open source platform to orchestrate and manage Docker containers is  used to coordinate the microservices in the PaaS layer. Kubernetes is extremely useful for the monitoring and scaling of services, and to ensure their reliability. The PaaS manages the needed micro-services using Kubernetes, in order, for example, to select the right end-point for the deployment of applications or services.  The Kubernetes solution is used in the PaaS layer as is provided by the community. 

The microservices that compose the PaaS layer are very heterogeneous in terms of development: some of them are developed ad hoc, some others were already available and used as they are, few others are significantly modified in order to implement new features within INDIGO. 

The language in which the INDIGO PaaS receives end user requests is TOSCA\cite{TOSCA}. TOSCA stands for Topology and Orchestration Specification for Cloud Applications. It is an OASIS specification for the interoperable description of applications and infrastructure cloud services, the relationships between parts of these services, and their operational behaviour. 

TOSCA has been selected as the language for describing applications, due to the wide-ranging adoption of this standard, and since it can be used as the orchestration language for both OpenNebula (through the IM) and OpenStack (through Heat).

\begin{figure}
  \centering
  \includegraphics[width=\textwidth]{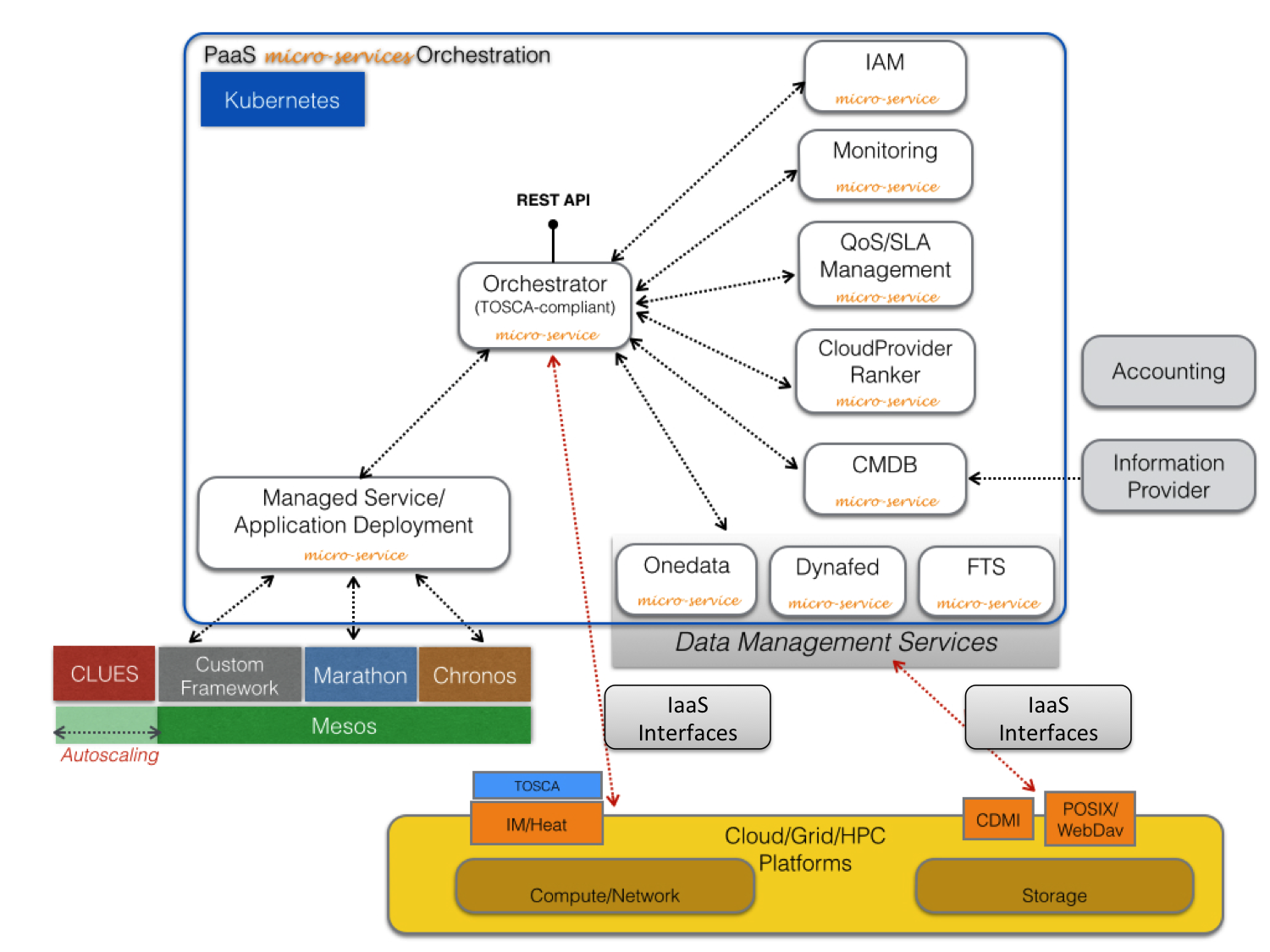}
  \caption{Architecture of the INDIGO Platform as a Service layer.}
  \label{fig:3}
\end{figure}

The released INDIGO PaaS layer (see Figure \ref{fig:3}) is able to provide automatic distribution of the application and/or services over a hybrid and heterogeneous set of IaaS infrastructures, on both private and public clouds.

The PaaS layer is able to accept a description of a complex set, or cluster, of services/applications by mean of TOSCA templates, and is able to provide the needed brokering features in order to find the best fitting resources. During this process, the PaaS layer is also able to evaluate data distribution, so that the resources requested by the users are chosen by the closeness to the storage services hosting the data requested by those specific applications/services.

\begin{figure}
  \centering
  \includegraphics[width=\textwidth]{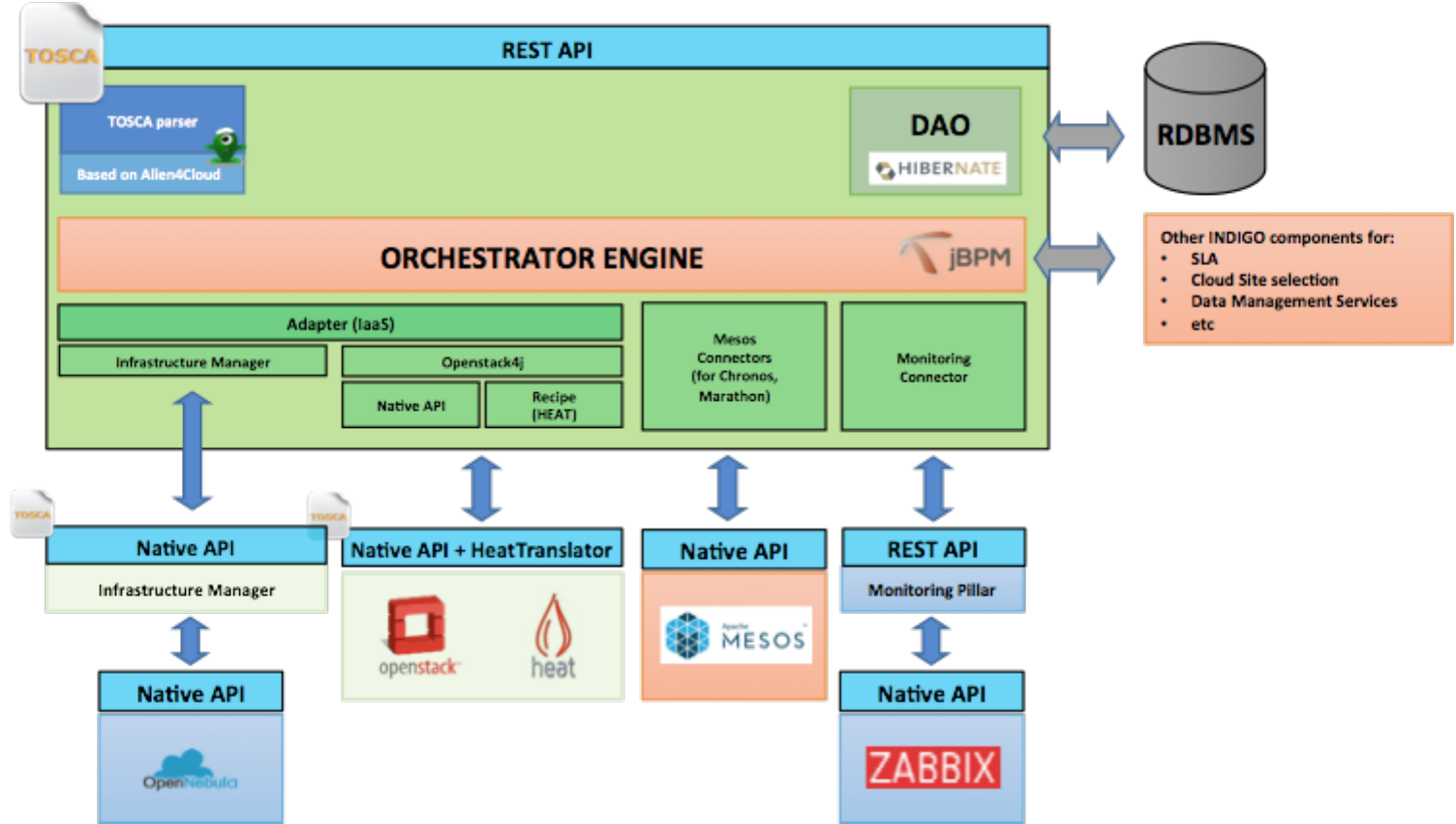}
  \caption{Architecture of the Orchestrator within the PaaS layer.}
  \label{fig:4}
\end{figure}

\subsubsection{The Orchestrator Engine}

The INDIGO PaaS {\bf Orchestrator}\footnote{https://github.com/indigo-dc/orchestrator} is a core component of the PaaS layer: it orchestrates the provisioning of virtualized compute and storage resources on Cloud Management Frameworks (like OpenStack and OpenNebula) and on Mesos clusters.

It receives the deployment requests, expressed through templates written in TOSCA, and deploys them on the best available cloud site. In order to select the best site, the Orchestrator implements a complex workflow: it gathers information about the SLAs signed by the providers and monitoring data about the availability of the compute and storage resources. Finally the Orchestrator asks the {\bf Cloud Provider Ranker}\footnote{https://github.com/indigo-dc/cloudproviderranker} to provide a ranked list of best cloud providers according an algorithm described below. Therefore the Orchestrator mission is to coordinate the deployment process over the IaaS platforms. See Figure \ref{fig:4} for an overview of the Orchestrator architecture.

The Orchestrator is based on developments already done in other publicly funded projects, such as the Italian PONs PRISMA\cite{PRISMA}, and Open City Platform\cite{OPENCITY}. During the INDIGO project, this component has been extended and enhanced to support the specific microservices building the INDIGO PaaS Layer. It delegates the actual deployment of resources to IM, OpenStack Heat or Mesos frameworks based on TOSCA templates and the ranked list of sites.

A very innovative component is the Cloud Provider Ranker. This is a standalone REST web service, which ranks cloud providers on the basis of rules defined per user/group/use case, with the aim of fully decoupling the ranking logic from the business logic of the Orchestrator.

It allows the consumers of the service (one or more orchestrators) to specify preferences on cloud providers. If some preferences have been specified for some providers, then they have absolute priority over any other provider information (like monitoring data). On the other hand, when preferences are not specified, for each provider the rank is calculated, by default, as the sum of SLA ranks and a combination of monitoring data, conveniently normalized with weights specified in the Ranker configuration file. Moreover, the ranking algorithm can be customized to the specific needs.

This is a completely new service, fully implemented within the INDIGO project; it is based on an open source tool, Drools\footnote{https://www.drools.org}, in order to reduce the needed development effort, and to simplify the long-term support.

As a summary, one of the main achievements of INDIGO in this respect is the implementation of TOSCA Templates on IaaS that do not support natively TOSCA, like Standard OpenStack, OpenNebula, or Public clouds (like Microsoft Azure, AWS, OTC,...) using the Infrastructure Manager.

In particular, using the PaaS Orchestrator and the TOSCA templates, the end user can exploit computational resources without knowledge about the IaaS details: indeed the TOSCA standard language ensures that the same template can be used to describe a virtual cluster on different cloud sites; then the Infrastructure Manager implements the TOSCA runtime for contacting the different cloud sites through their native APIs. The provisioning and configuration of the IaaS resources is therefore accomplished in a completely transparent way for the end user. The same approach is used also for submitting dockerized applications and services to a Mesos cluster (and its frameworks Marathon and Chronos): the user can describe his request in a TOSCA template and the Orchestrator provides the TOSCA runtime for contacting the Mesos master node, submitting the request and monitoring its status on behalf of the user as detailed in the next section.

\subsubsection{High-level geographical applications/service deployment}


INDIGO has developed the tools and services to provide a solution for orchestrating Docker containers for both applications (job-like execution) and long running services. {\bf Mesos/Marathon/Chronos}\footnote{https://github.com/indigo-dc/mesos-cluster} is used to manage the deployment of services and applications (MSA service). 

From a resource perspective Mesos is a cluster management tool: it pools several resource centers to be centrally managed as single unit; from an application perspective, Mesos is a scheduler: it dispatches workloads to consume pooled resources, scaling up to thousands of nodes. 

Mesos is fault tolerant, as it is possible to replicate the master process. The INDIGO PaaS uses also {\bf Marathon} and {\bf Chronos}. Marathon is used to deploy, monitor and scale Long-running services, ensuring that they are always up and running. Chronos is used to run user applications (jobs) taking care of fetching input data, handling dependencies among jobs or rescheduling failed jobs.

Therefore the submission of jobs uses an approach very similar to a batch system, exploiting resources where there are, without even knowing about the details. This includes deployment on multiple IaaS both private and public, hiding to the end users the complexity of the distributed resources.

It is also possible using the service {\bf CLUES}\footnote{https://github.com/indigo-dc/clues-indigo} (Cluster Energy Saving) to auto-scale (up \& down) depending on the load, on several types of clusters: Mesos, SLURM, PBS, HTCondor, etc. CLUES is an elasticity manager system for HPC clusters and Cloud infrastructures that features the ability to power on/deploy working nodes as needed (depending on the job workload of the cluster) and to power off/terminate them when they are no longer needed.


The sustainability of our PaaS layer relies heavily on the fact that we have used Open Source frameworks when already available.

\subsubsection{Data Management Services}

The goal of the data management developments in INDIGO has been pushing forward the state of the art concerning unified data access over heterogeneous infrastructures. Among he features demanded by the use cases are High-performance data access, migration and replica management. Supporting such features requires at the user level a flexible security framework based on tokens and Access Control Lists (ACLs). 

Data management services developed in INDIGO are based on three open source components: Onedata\cite{ONEDATA}, DynaFed\cite{DYNAFED} and FTS3\cite{FTS}.

INDIGO has invested a substantial effort in the development of {\bf Onedata}\footnote{https://github.com/indigo-dc/onedata}, this is a global data management system aiming to provide easy access to distributed storage resources. The final goal is supporting a  wide  range  of  use  cases  from  personal  data  management  to  data-intensive  scientific  computations. 

Support for federation in  Onedata can be achieved by the possibility of establishing a distributed provider registry, where various infrastructures can setup their own provider registry and build trust relationship between these instances, allowing users from various platforms to share their data transparently.

Onedata  provides  an  easy  to  use  Graphical  User  Interface  for  managing  storage  Spaces,  with customizable   access   control   rights   on   entire   data   sets   or   single   files   to   particular   users   or groups. 

The INDIGO PaaS Orchestrator integrates a plugin for interacting with the Onedata services providing advanced capabilities of data location aware scheduling. Combining the information about the distribution of the compute resources and the data providers with the data requirements specified by the user, the Orchestrator is able to schedule the processing jobs to the computing center nearest to the data. A prototype based on Onedata has been implemented and demonstrated for some use-cases, e.g. the LifeWatch AlgaeBloom (See Table \ref{tab:req}).

Using Onedata is possible to integrate already available storage services in
the INDIGO Platform exploiting the data stored in external infrastructures. This is the case of the data stored in WLCG by the CMS experiment from LHC. In this case the INDIGO PaaS provided the services needed in order to deal with authentication and
autorization (Token Translation)


\section{Interfacing with the users}
\label{sec:users}

Users typically do not access the PaaS core components directly. They instead often use Graphical User Interfaces or simpler APIs. A user authenticated on the INDIGO Platform can access and customize a rich set of TOSCA-compliant templates through a GUI-based portlet.

The INDIGO repository provides a catalogue of pre-configured TOSCA templates to be used for the deployment of a wide range of applications and services, customizable with different requirements of scalability, reliability and performance. In these templates a user can choose between two different examples of generic scenarios:
 
\begin{enumerate}
\item Deploy a customized virtual infrastructure starting from a TOSCA template that has been imported, or built from scratch: The user will be able to access the deployed customized virtual infrastructure and run, administer and manage applications running on it.

\item Deploy a service/application whose lifecycle will be directly managed by the PaaS platform: in this case the user will be returned the list of endpoints to access the deployed services.
\end{enumerate}

APIs for accessing the INDIGO PaaS layer are available, they allow for an easy integration of the PaaS features inside Portals, Desktop Applications and Mobile Apps. The final release of INDIGO-DataCloud software includes a large set of components to facilitate the development of Science Gateways and desktop/mobile applications, big data analytics and scientific workflows. The components directly related to end-user interfaces are:

\begin{itemize}
\item The INDIGO FutureGateway (FG) framework, used to build powerful, customized, easy to use, science gateway environment and front-ends, on top of the INDIGO-DataCloud PaaS layer and integrated with data management services. The FG provides many capabilities, including: 
\item The FG API server, used to integrate third-party science gateways; the FG Liferay Portal, containing base portlets for the authentication, authorization and administration of running applications and deployments; 
\item Customizable Application Portlets, for user-friendly specification of the parameters used by TOSCA templates;  
\item A workflows monitoring portlet, used for monitoring task execution via integrated workflow systems, described below. 
\item An Open Mobile Toolkit as well as application templates for Android and iOS, simplifying the creation of mobile apps making use of the FG API Server. 
\item Support for scientific workflows, where the INDIGO components: 
\begin{itemize}
\item Provide dynamic scalable services in a Workflows as a Service model;
\item Implement modules and components enabling the usage of the PaaS layer (via FG API Server) for the main scientific workflow engines deployed by user communities (such as Kepler, Ophidia, Taverna,Pegasus);  
\item Support a two-level (coarse and fine grained) workflow orchestration, essential for complex, distributed experiments involving (among others) parallel data analysis tasks on large volumes of scientific data.    
\end{itemize}
\item Key extensions to the Ophidia big data analytics framework (allowing to process, transform and manipulate array-based data in scientific contexts), providing many new functionalities, including a set of new operators related to data import regarding heterogeneous data formats (e.g. SAC, FITS), a new OpenIDConnect interface and new workflow interface extensions. 

\item Enhancements of the jSAGA library through a "Resource Management API", complementing the standard Job/Data Management API. This allows to acquire and manage resources (compute, storage, network) and enables the wrapping of underlying technologies (cloud, pilot jobs, grid, etc.) by means of a single API, supporting asynchronous mode (task), timeout management, notification (metrics) and security context forwarding. 

\item Command-line clients for the PaaS layer to provide an easy way for users to interact with the Orchestrator or with WATTS:
\begin{itemize}
\item {\bf Orchent}\footnote{https://github.com/indigo-dc/orchent}: a command-line application to manage deployments and their resources on the INDIGO-DataCloud Orchestrator;
\item {\bf Wattson}\footnote{https://github.com/indigo-dc/wattson}: a command-line client for the INDIGO Token Translation Service.
\end{itemize}

\end{itemize}

\begin{figure}
  \centering
  \includegraphics[width=\textwidth]{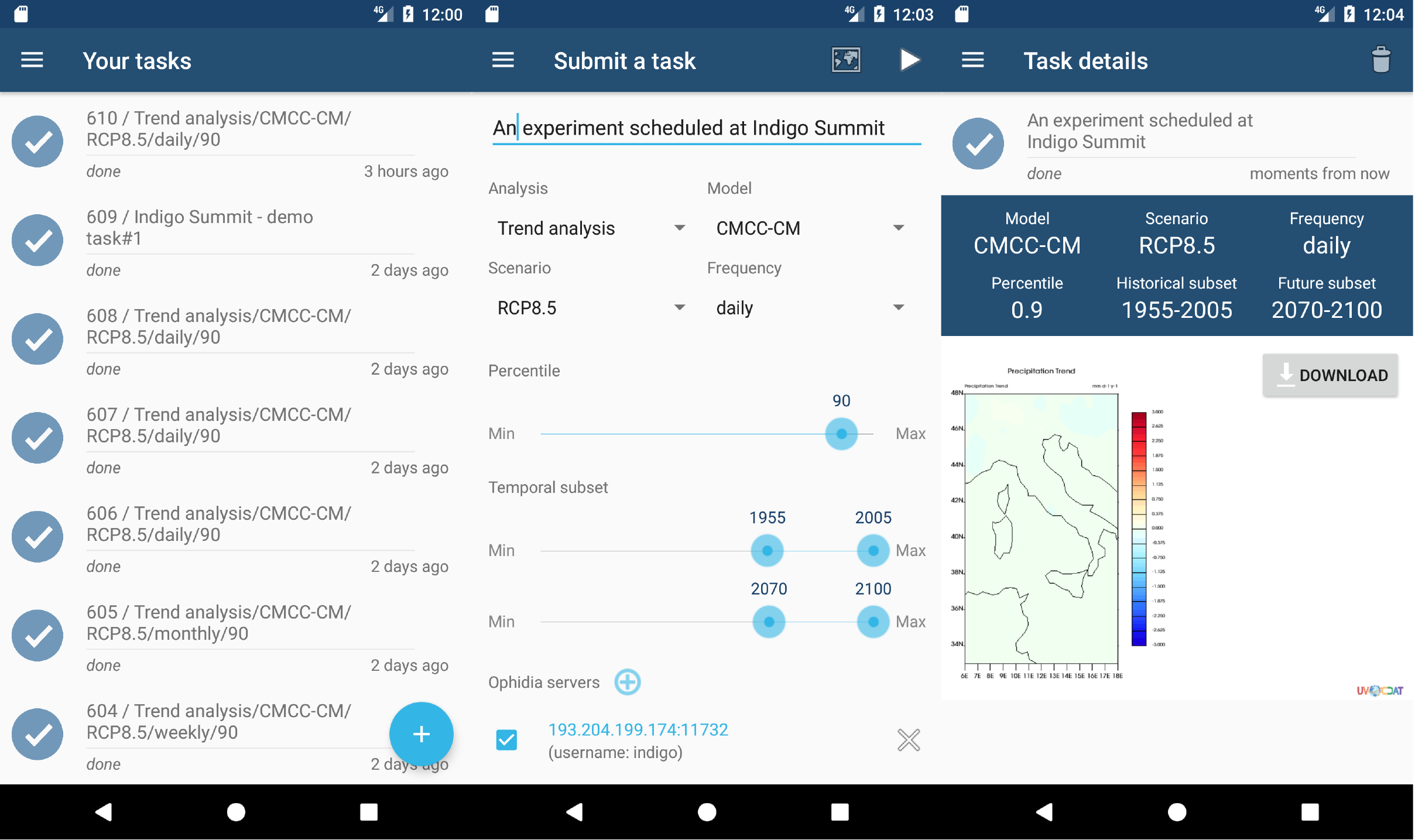}
  \caption{Task submission and visualization of results on the mobile platform for ENES}
  \label{fig:18}
\end{figure}

INDIGO has provided the tools for a simple and effective end user experience, both for software developers and for researchers running the applications. In Figure \ref{fig:18} we show how to launch an application using a mobile platform developed in the project for the climate change application ENES.

The ENES end-user  starts  the  app  and  needs  to  authenticate  and  authorize  using  IAM  service. The  apps  request  to  have  an  access  to  id, e-mail  and  offline  access,  which  will  be  required to refresh  the  existing  tokens  during  the mobile  app  lifecycle.  After  the  user  gives  the  permissions and logs in, the user  will  see  the  list  with  scheduled  analysis  if  available.  

The  mobile  app  is  a  handy  interface  for  scheduling  new tasks  as  well.  The submitting  form requires that the user selects  model,  scenario,  frequency,  percentile  and  nodes  to  run  the  analysis.  After  the user  provides the necessary  inputs,  the  app  sends  the  request  to  the  FutureGateway  server using  its  API. The  user  is then  able  to  monitor  the  status  of  the  analysis.  If  the  task  is  done, the user  is  able  to  see  and  download  results  as  PNG  files  illustrates  predicted  climate  changes on Earth.  Finally  the  user  can  remove  useless  or  aborted  tasks.

\section{Software lifecycle management}
\label{sec:release}

\begin{figure}[h]
  \centering
  \includegraphics[width=\textwidth]{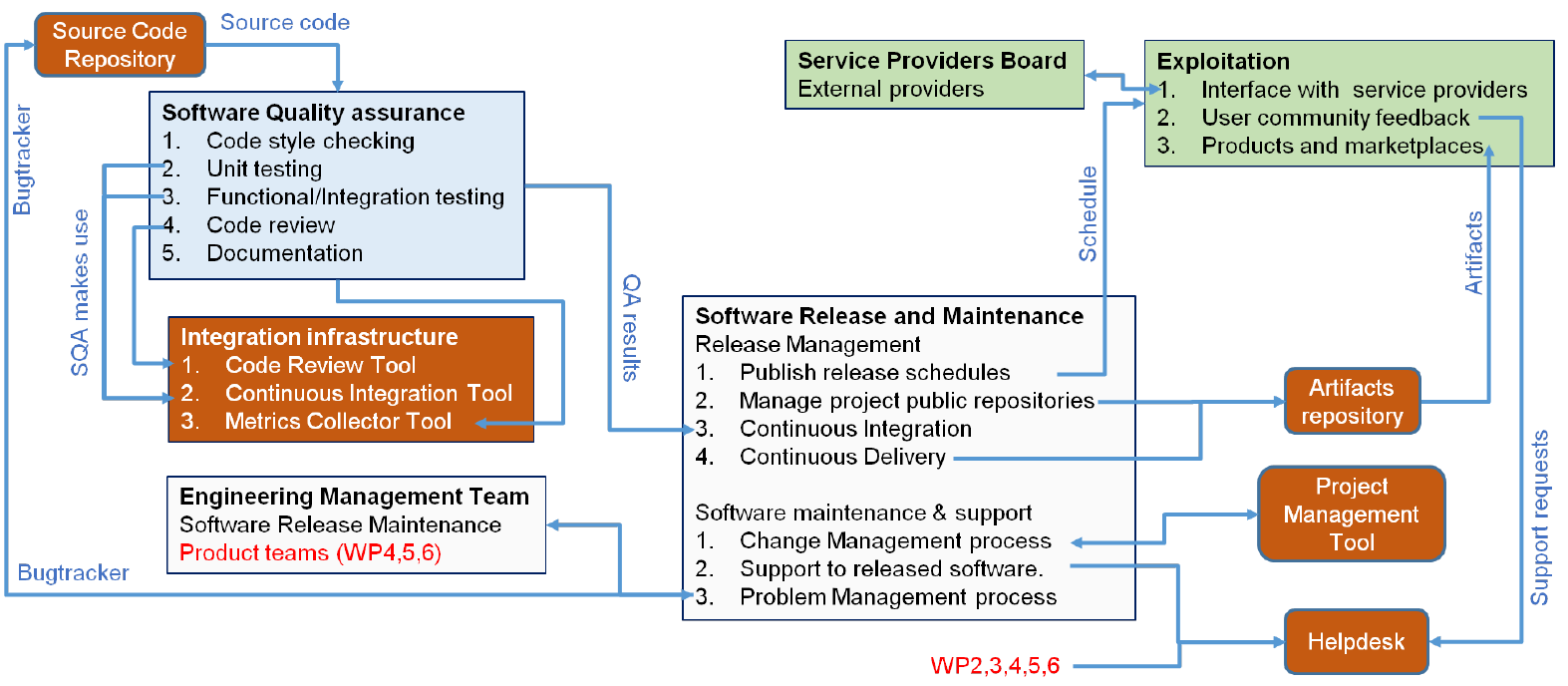}
  \caption{Software lifecycle, release, maintenance and exploitation interdependencies.}
  \label{fig:6}
\end{figure}

The software lifecycle process in INDIGO has been supported by a continuous improvement process that encompassed the software quality assurance (SQA), the software
release and maintenance, the deployment of pilot infrastructures for software integration and testing,
and, lastly, the exploitation activities and support services. In Figure \ref{fig:6} we
depict the interdependencies between the different processes, together with the 
services involved at each stage. Appendix \ref{appendix:tools} describes the tools and
services that were required for the implementation of the software lifecycle process. 

The quality requirements \cite{SQA}, that drive the software lifecycle process, define
the minimum set of criteria that the software developed in INDIGO has to comply with. 
The requirements are met for each change in the codebase, thus the production version
of a given software component is permanently in a workable status, protected from 
incoming changes that do not adhere to the SQA criteria. The continuous evaluation of 
the SQA requirements is only possible through the aid of automation, achieved in INDIGO
through the progressive adoption of DevOps practices.

In the next sections we will describe the DevOps approaches being adopted and the upstream
contributions included in the official distributions of external open source projects.

\subsection{DevOps approach in INDIGO}

Progressive levels of automation were being adopted throughout
the different phases of the INDIGO-DataCloud project software development and
delivery processes. This evolution was intentionally marked by the commitment
to DevOps principles \cite{devops}. Starting with a continuous integration (CI)
approach, achieved already in the early stages of the project, the second part
of the project was devoted to the establishment of the next natural step in the
DevOps practices: the continuous delivery (CD).

\subsubsection{Services for continuous integration and SQA}

\begin{figure}
  \centering
  \includegraphics[width=\textwidth]{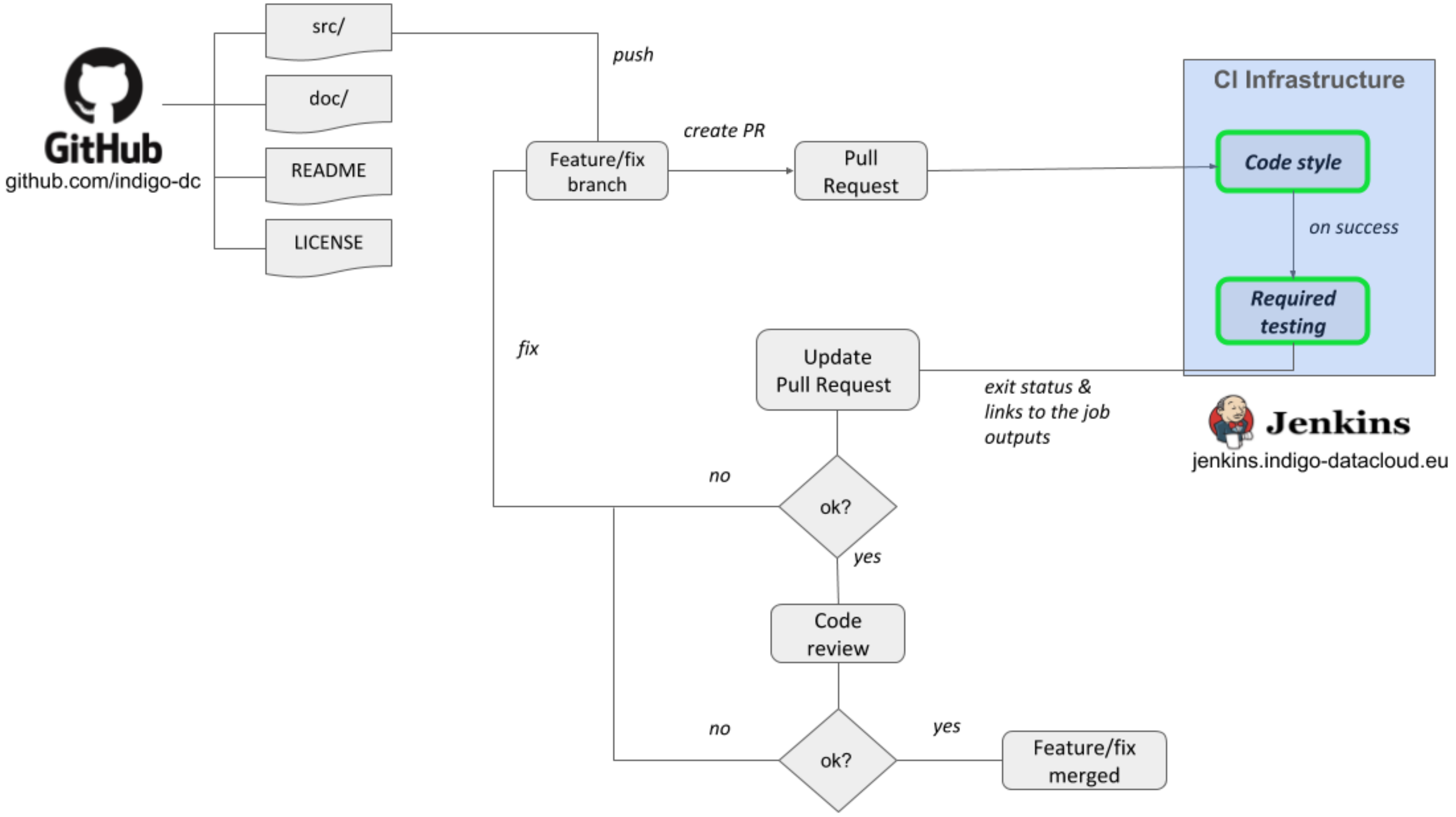}
  \caption{Continuous Integration workflow followed by new feature additions in the production codebase.}
  \label{fig:8}
\end{figure}

The INDIGO-DataCloud CI process is schematically shown in Figure \ref{fig:8} is
explained below. The process, in its different steps, reflects some of the main
and important achievements of the software integration team.

\begin{itemize}
    \item New features are developed independently from the production version
        in \textit{feature branches}. In order to test and review the new change,
        a pull request (PR) is created in GitHub. The PR creation marks the start
        of the automated validation process through the execution of the SQA jobs
        in the CI infrastructure (Jenkins).
    \item The SQA jobs perform the adherence of the code to a style standard and 
        calculate unit and functional test coverage. Other checks are executed at
        this stage for security static analysis and metrics gathering.
    \item The results of the several SQA jobs automatically executed in Jenkins
        are notified back to GitHub, updating the PR with the exit status and the
        links to the output logs.

    \item On successful completion of the SQA tests, the code review is the last
        step before the source code is merged in the production version. The 
        GitHub PR provides a place for discussion, open to collaboration, where
        the developers and/or external experts analyze the results of the SQA jobs and
        discuss any relevant aspect of the change (internal to the code or in
        terms of the goals or applicability).
    \item Once peer-reviewed, the change is merged and becomes ready for 
        integration and later release.
\end{itemize}

\subsubsection{Continuous delivery}
Continuous delivery adds, on top of the CI approach described above, a seamless
manufacturing of software packages ready to be deployed into production.

In the INDIGO-DataCloud scenario, the continuous delivery adoption translates
into the definition of pipelines. A pipeline is a serial execution of tasks
that encompasses in the first place the SQA jobs (CI phase) and adds as the
second part (CD phase) the building and deployment testing of the software
packages created. The pipeline only succeeds if each task is run to completion,
otherwise the process is stopped and set as a build failure.

\subsubsection{DevOps adoption from user communities}

The experience gathered throughout the project with regards to the adoption of
different DevOps practices is not only useful and suitable for the software
related to the core services in the INDIGO-DataCloud solution, but also
applicable to the development and distribution of the applications coming from
the user communities.

The novelty introduced, showcased in Appendix \ref{appendix:devops:users}, is the 
validation of the user application by comparing the execution results with a
set of reference outputs. Thus this pipeline implementation goes a step forward, 
with respect to the former DevOps approaches, as the application execution is 
tested before the new version is released.

\subsection{INDIGO upstream software contributions}

The INDIGO software solution encompasses 
not only products implemented from scratch within the project but also external services
adopted from open source initiatives. These latter set of products were actively developed
to enhance their functionality to match the INDIGO project's objectives, but at the same
time, aiming to be considered as upstream contributions. Thus, multiple contributions 
developed by INDIGO have been pushed and accepted in the official distributions of major
open source projects such as OpenStack, OpenNebula and OpenID Connect. Appendix 
\ref{appendix:upstream} lists the software projects and products being contributed by 
INDIGO-DataCloud.


\section{Examples of implementation towards research communities}
\label{sec:examples}

In what follows we try to provide some basic information that may be useful for promoting the use of INDIGO solutions towards the Research Communities. Based on the described architecture we will introduce the basic ideas on how to develop, deploy and support applications in the Cloud framework, exploiting the different service layers, and introducing generic examples that may make easier the use of INDIGO solutions.

\subsection{Understanding the services of the Cloud Computing framework}

\begin{figure}
  \centering
  \includegraphics[width=\textwidth]{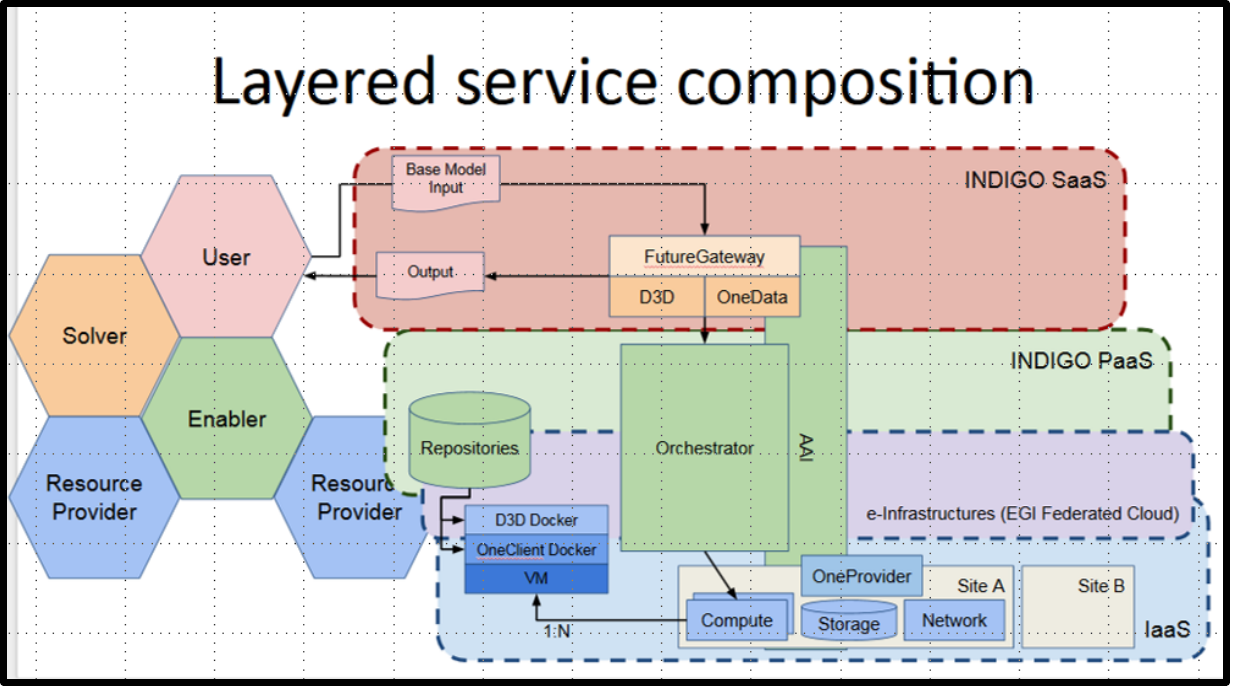}
  \caption{Service composition in the Service Oriented Architecture}
  \label{fig:11}
\end{figure}

Figure \ref{fig:11} provides the description of how an application can be built using a service oriented architecture in the Cloud, using INDIGO solutions. 

This layered scheme includes different elements, that are managed by different actors, that must be minimally understood in order to design, develop, test, deploy and put in production an application.

The lowest layer, Infrastructure as a Service, provides a way to access to the basic resources that the application will use: computing, storage, network, etc. These resources are physically in a site, typically a computing center, either in a research centre, or in a cloud provider (for example commercial cloud services), and are handled by the system managers at those sites, that install a IaaS solution compatible with the INDIGO software stack (e.g. OpenStack, OpenNebula, Google Compute Engine, Microsoft Azure, Amazon EC2).

By accessing through a web interface such as Horizon for OpenStack, a user that is granted access to a pool of resources at a site can launch a virtual machine, for example a server with 2 cores, 4GB RAM, 100GB of storage and with a certain Linux flavour installed. 

The user can then get access in console mode this machine, using for instance ssh protocol. Once logged in, the user can execute a simple script, or can install a web server, etc. When the work is finished the machine can be stopped by the user, liberating so the resources. This is a very basic mode of accessing Cloud services, which shares analogies with the usual access to classical computing services, like for example any remote server or a cluster.

A different way to interact with IaaS services is to use the existing APIs to manage the resources using the web services protocol. Such invocation of services can be made from any program or application, for example from a python script, and even through a web interface. 

Many applications require the setup, launching and interconnection of several (IaaS) services implemented in different virtual machines, and managed under a single control, as a Platform. The Platform as a Service (PaaS) layer enables this orchestration of IaaS services, and in the case of a Federated Cloud they might even be located in different sites.

For example, an Apache Web Server may be launched in site 1, with the purpose of displaying the output of a simulation running in a cluster launched on demand in site 2. The Apache Web server may be better supported with a pool of resources using another cloud-oriented solution such as Marathon/Mesos. Launching an application in this context requires identifying the available resources and launching them via the IaaS services. INDIGO is supporting the TOSCA standard to prepare a template that can be used to automatize this selection and orchestration of services.

\subsection{Building and executing applications using INDIGO solutions}

In what follows we present below several simple examples of basic, but generic, applications exploiting INDIGO solutions. 

\subsubsection{Executing containers on HPC systems}

\begin{figure}
  \centering
  \includegraphics[width=\textwidth]{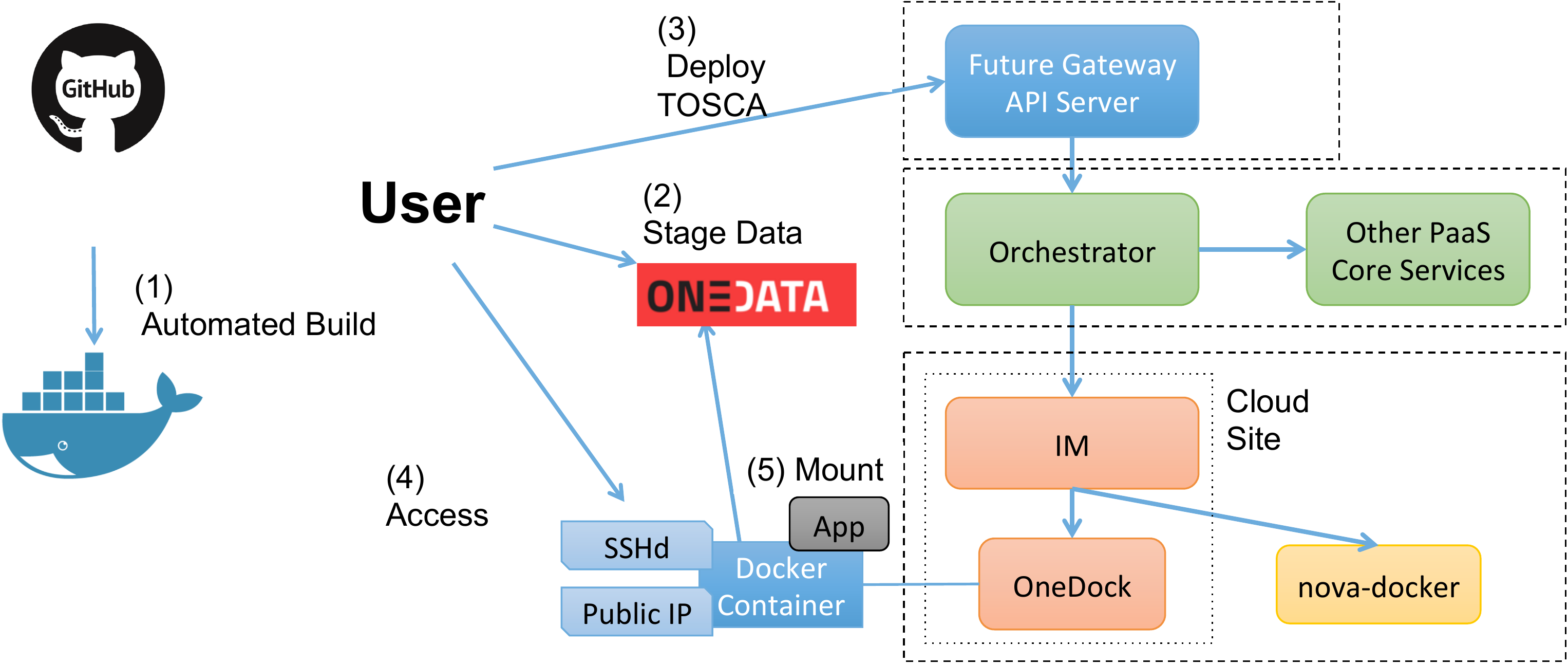}
  \caption{Basic execution of containers using INDIGO solutions}
  \label{fig:12}
\end{figure}

The first generic example is how to build an application encapsulated as a container and how to executed it in an HPC system. This basic example of using INDIGO solutions is shown in Figure \ref{fig:12}. A user can create a container using a conventional Dockerfile which describes the steps required to create the Docker image. The process can be fully automatized using GitHub and Docker Hub in such a way that a change in the Dockerfile, immediately triggers a rebuilt of the application container.

INDIGO provides the udocker tool to enable execution of application containers in batch systems. The end-user can download the udocker Python script from GitHub or can send it with the batch job. Once executed for the first time it setups itself in the user home directory. udocker provides a Docker like command line interface with which the user can pull, import or load Docker containers and then execute them using a chroot-like environment. The software within the container must not require privileges during execution as it will be executed under the user that invokes udocker.

This is also a good solution for research communities that want to migrate towards a cloud-based framework using containers, but keep exploiting resources like grid-enabled clusters or even supercomputers. 

udocker is used by the Case Studies on Structural Biology (Powerfit and Disvis) exploiting grid resources, on Phenomenology in Particle Physics, and recently for Lattice QCD on supercomputers. Also, the TRUFA genomic pipeline exploits this solution, and it is being extended to similar applications in the area that require the integration of legacy libraries.

\subsubsection{Executing containers on the Cloud}

The second example is how to build an application encapsulated as a container and launch it in the Cloud, from a web interface, using the INDIGO solutions FutureGateway and PaaS Orchestrator

This second example, compared to the first one, shows the evolution required to move an application to the Cloud arena: the application must be encapsulated into a container, as before, but to launch this container the cloud resources must be allocated, the user must authenticate and get the access granted. If different services are required, they must be orchestrated.

The way to express these requirements, using an open standard, is a TOSCA template. FutureGateway offers a user-friendly web-based interface to customize the TOSCA template, authenticate the user, select the container to be executed, interact with the Orchestrator to allocate the required cloud resources and launch the application. As in the first example the container can be created using a dockerfile. Automation in this step can be achieved using GitHub and DockerHub.

Using the Future Gateway portal or the command line tool Orchent, the user can submit a TOSCA template to the Orchestrator, which in turn will request and allocate the resources at the IaaS level by asking the Infrastructure Manager to do so.

The user may wish to connect to the container that has been launched via the orchestrator using the {\tt ssh} command. Once in the container it is posible to mount remote data repositories or stage the output data using Onedata, available at the IaaS layer.

\subsection{Building advanced applications using INDIGO solutions}

In this section we present several simplified schemes corresponding to different applications already implemented, with the idea that they can be more easily used as a guide to configure new applications.  

The key, as stated before, is the composition of the template, written using the TOSCA language. The template should specify the image of the application to be used, as a container, using docker technology. We also need to specify the resources (CPUs, storage, memory, network ports) required to support the execution. The parameters required to configure INDIGO services used like, for example, Onedata, Mesos/Marathon or other additional cloud services need to be specified as well.

Examples of TOSCA templates can be found at https://github.com/indigo-dc/tosca-templates. FutureGateway offers a friendly way to handle the TOSCA templates to launch the applications.

\subsubsection{Deployment of a Digital Repository}

A first example is the deployment, as a SaaS solution, of a digital repository. The scheme is presented in the Figure \ref{fig:13} below. The specific template for this application is available for reuse in the github repository of the project.

The repository manager, controlling the application, uses the FutureGateway to configure the application, based on the ZENODO software, that can be automatically scaled up and ensure its high availability using Cloud resources as needed, and also enabling the authentication and authorization mechanism  for their research community, DARIAH, based on the INDIGO solution IAM. 

All these details are transparent to the final user, who accesses the repository directly through its web interface, and benefits of the enhanced scalability and availability.

\begin{figure}
  \centering
  \includegraphics[width=\textwidth]{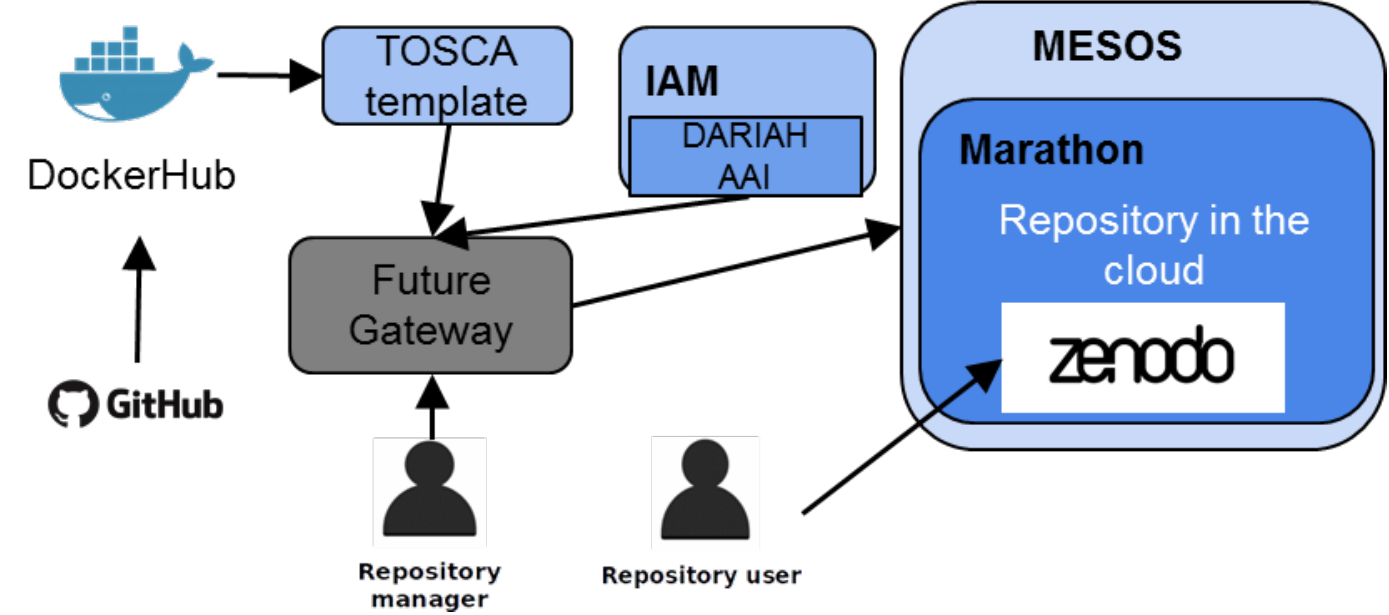}
  \caption{Deployment of a digital repository using INDIGO solutions}
  \label{fig:13}
\end{figure}

\subsubsection{Launching a Virtual Elastic Cluster for Data Intensive Applications}

A second example is the launch of a Virtual Elastic Cluster to support a data intensive system.The scheme is presented in Figure \ref{fig:14} below. 

The specific template for this advanced application is available for reuse in the github repository of the project.

Galaxy is an open source, web-based platform for data intensive biomedical research. This application deploys a Galaxy instance provider platform, allowing to fully customize each virtual instance through a user-friendly web interface, ready to be used by life scientists and bioinformaticians.

The front-end that will be in charge of managing the cluster elasticity can use a specified LRMS (selected among torque, sge, slurm and condor) workload.

All these details are transparent to the final user, the researcher, who accesses the Galaxy instance directly through its web interface, and benefits of the enhanced scalability and availability.

This complex template includes the configuration of the distributed storage based in Onedata, the use of the encrypted files via LUKS, the deployment of elastic clusters using another INDIGO solution, CLUES, and the integration of the Authentication and Authorization mechanism, very relevant for this application area, using IAM.  

\begin{figure}
  \centering
  \includegraphics[width=\textwidth]{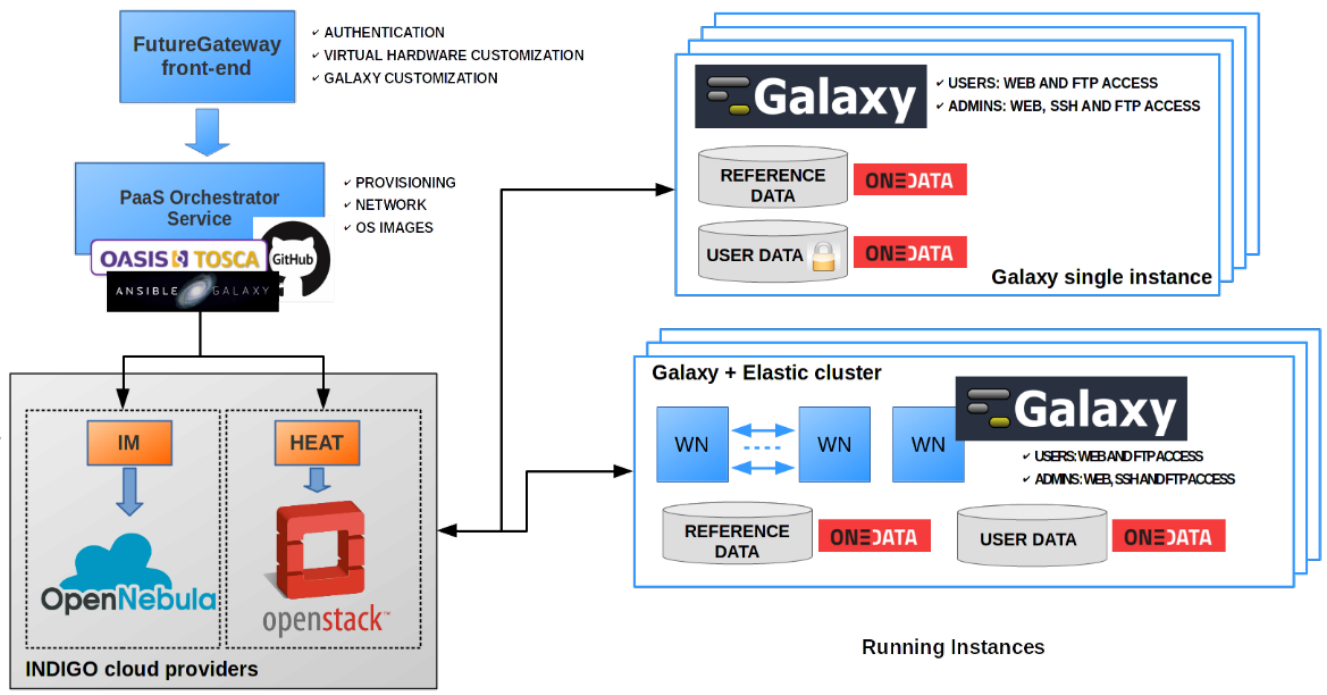}
  \caption{Launching of a Virtual Elastic Cluster using INDIGO solutions}
  \label{fig:14}
\end{figure}

\section{Conclusions}
\label{sec:conclusions}

Thanks to the new common solutions developed by the INDIGO project, teams of first-line researchers in Europe are using public and private Cloud resources to get new results in Physics, Biology, Astronomy, Medicine, Humanities and other disciplines. 

INDIGO-developed solutions that have for instance enabled new advances in understanding how the basic blocks of matter (quarks) interact, using supercomputers, how new molecules involved in life work, using GPUs, or how complex new repositories to preserve and consult digital heritage can be easily built. The variety of the requirements coming from these so diverse user communities proves that the modular INDIGO platform, consisting of several state-of-the-art, production-level services, is flexible and general enough to be applied to all of them with the same ease of use and efficiency. 

These services allow to federate hybrid resources, to easily write, port and run scientific applications to the cloud. They are all freely downloadable as open source components, and are already being integrated into many scientific applications, namely: 

\begin{itemize}
\item High-energy physics: the creation of complex clusters deployed on several Cloud infrastructures is automated, in order to perform simulation and analysis of physics data for large experiments. 
\item Lifewatch: parameters from a water quality model in a reservoir are calibrated, using automated multiple simulations. 
\item Digital libraries: multiple libraries can easily access a cloud environment under central coordination but uploading and managing their own collections of digital objects. This allows them to consistently keep control of their collections and to certify their quality. 
\item Elixir: Galaxy, a tool often used in many life science research environments, is automatically configured, deployed on the Cloud and used to process data through a user-friendly interface. 
\item Theoretical HEP physics: the MasterCode software, used in theoretical physics, adopts INDIGO tools to run applications on Grids, Clouds and on HPC systems with an efficient, simple-to-use, consistent interface. 
\item In DARIAH, a pan-european social and technical infrastructure for arts and humanities, the deployment of a self-managed, auto-scalable Zenodo-based  repository in the cloud is automated. 
\item Climate change: distributed, parallel data analysis in the context of the Earth System Grid Federation (ESGF) infrastructure is performed through software deployed on HPC and cloud environments in Europe and in the US. 
\item Image analysis: in the context of EuroBioImaging, a distributed infrastructure for microscopy, molecular and medical imaging, INDIGO components are used to perform automatic and scalable analysis of bone density. 
\item Astronomical data archives: big data consisting of images collected by telescopes are automatically distributed and accessed via INDIGO tools. 
\end{itemize}

The same solutions are also being explored by industry, to provide innovative services to EU companies: for example, modelling water reservoirs integrating satellite information, improving security in cyberspace, or assisting doctors in diagnostics through medical images. INDIGO solutions are also being intensively tested in other projects, such as HelixNebula ScienceCloud. 

INDIGO services are fundamental for the implementation of the EOSC. In particular, many INDIGO components are included in the unified service catalogue provided by the project EOSC-hub \cite{EOSC-HUB}, that will put in place the basic layout for the European Open Science Cloud. Two additional Horizon 2020 projects were also approved ({\tt DEEP Hybrid DataCloud} and {\tt eXtreme DataCloud}), that will continue to develop and enhance INDIGO components. 

The outcomes of INDIGO-DataCloud will persist, and also be extended, after the end of the project in the framework of the {\sl INDIGO Software Collaboration agreement}. This Collaboration shall be continued without financial support from the European Union. It is open to new initiatives and partners willing to contribute, extend or maintain the INDIGO-DataCloud software components.

\section*{Acknowledgments}
INDIGO-Datacloud has been funded by the European Commision H2020 research and innovation program under grant agreement RIA 653549. 

\bibliography{references}
\bibliographystyle{unsrt}

\clearpage
\newpage
\begin{appendices}

\section{Contribution to Open Source software projects}
\label{appendix:upstream}

Here follows the list of software developed in the framework of INDIGO-Datacloud that has been contributed upstream to the Open Source community.

\begin{itemize}
    \item OpenStack (https://www.openstack.org)
    \begin{itemize}
        \item Changes/contribution done already merged upstream
            \begin{itemize}
                \item Nova Docker
                \item Heat Translator (INDIGO-DataCloud is 3rd overall contributor and core developer)
                \item TOSCA parser (INDIGO-DataCloud is 2nd overall contributor and core developer)
                \item OpenID Connect CLI support
                \item OOI: OCCI implementation for OpenStack
            \end{itemize}
        \item Changes/contribution under discussion to be merged upstream
            OpenStack Preemptible Instances support (extensions)
    \end{itemize}
    \item OpenNebula
        \begin{itemize}
            \item Changes/contribution done already merged upstream
            \begin{itemize}
                \item ONEDock
            \end{itemize}
        \end{itemize}
    \item Changes/contribution done already merged upstream for:
        \begin{itemize}
            \item Infrastructure Manager (http://www.grycap.upv.es/im/index.php)
            \item CLUES (http://www.grycap.upv.es/clues/eng/index.php)
            \item Onedata (https://onedata.org) 
            \item Apache Libcloud (https://github.com/apache/libcloud)
            \item Kepler Workflow Manager (https://kepler-project.org/)
            \item TOSCA adaptor for JSAGA (http://software.in2p3.fr/jsaga/dev/) 
            \item CDMI and QoS extensions for dCache (https://www.dcache.org) 
            \item Workflow interface extensions for Ophidia (http://ophidia.cmcc.it) 
            \item OpenID Connect Java implementation for dCache (https://www.dcache.org) 
            \item MitreID (https://mitreid.org/) and OpenID Connect (http://openid.net/connect/) libraries
            \item FutureGateway (https://www.catania-science-gateways.it/)
        \end{itemize}
\end{itemize}

\section{Tools and services involved in the software lifecycle}
\label{appendix:tools}

Figure \ref{fig:7} showcases the tools and services used for the development and distribution
of the INDIGO-DataCloud software:

\begin{figure}[h]
  \centering
  \includegraphics[width=\textwidth]{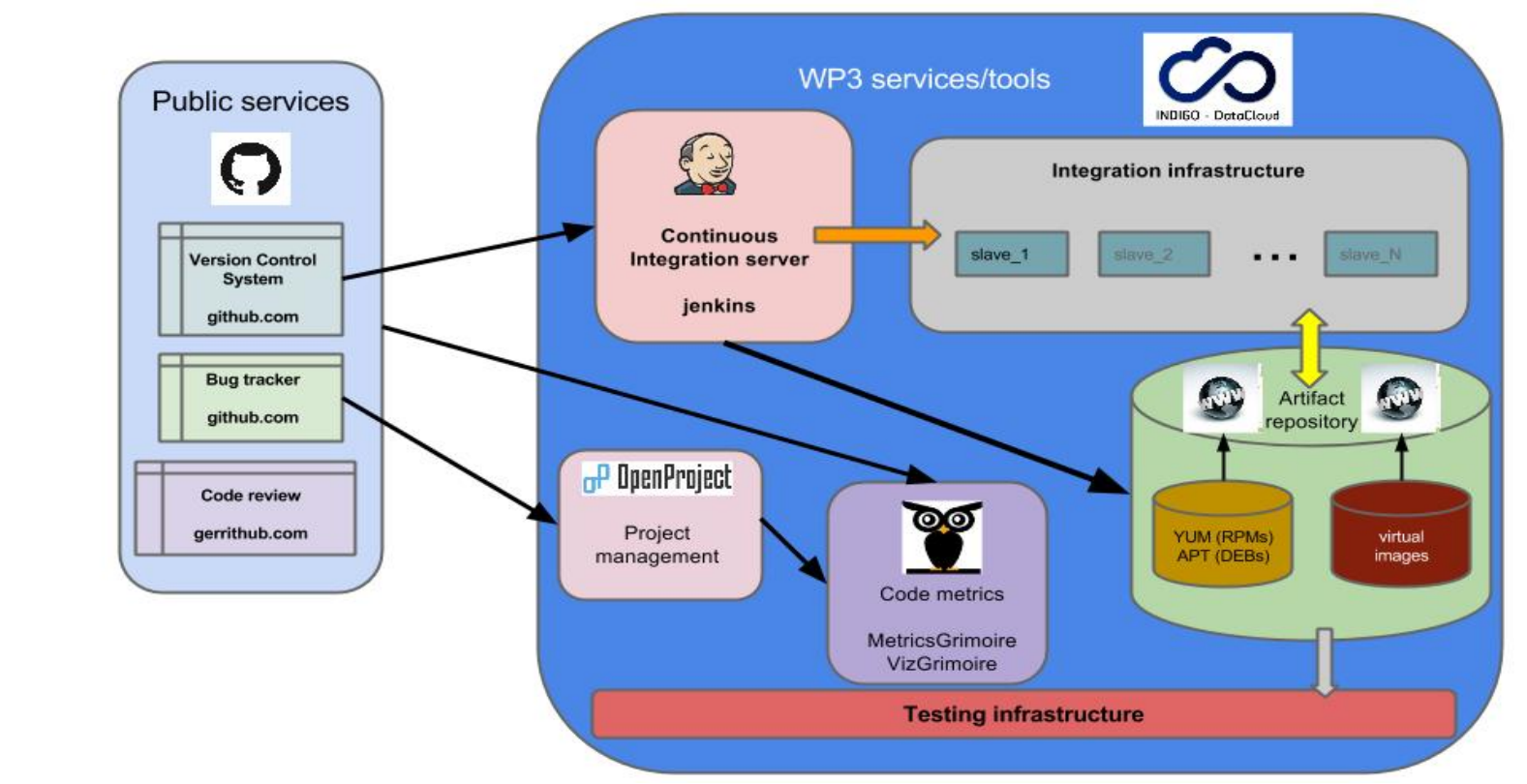}
  \caption{Tools and services used to support the software lifecycle process.}
  \label{fig:7}
\end{figure}

\begin{itemize}
    \item Project management service using {\bf openproject.org}: It provides
        tools such as an issue tracker, wiki, a placeholder for documents and a
        project management timeline.

    \item Source code is publicly available, housed externally
        in GitHub repositories, increasing so the visibility and simplifying the path to exploitation beyond the project lifetime. The INDIGO-DataCloud software is released under the Apache 2.0 software license \cite{apache-license}.

    \item Continuous Integration service using {\bf Jenkins}: Service to
        automate the building, testing and packaging, where
        applicable. Testing includes the style compliance and
        estimation of the unit and functional
        test coverage of the software components.

    \item Artifact repositories for RedHat and Debian
        packages \cite{indigo-package-repo} and virtual -- Docker -- images
        \cite{indigo-dockerhub}.

    \item Code review service using GitHub: Source code
        review is one integral part of the SQA
        as it appears as the last step in the change
        verification process. This service facilitates the code review
        process, recording the comments and
        allowing the reviewer to verify the candidate change before
        being merged into the production version.

    \item Issue tracking
        using GitHub Issues: Service to track issues, new features and bugs of
        INDIGO-DataCloud software components.

    \item Release notes, installation and configuration guides, user and
        development manuals are made available on {\bf GitBook}
        \cite{indigo-gitbook}.

    \item Code metrics services using {\bf Grimoire}: To collect and visualize
        several metrics about the software components.

    \item Integration infrastructure: this infrastructure is composed of
        computing resources to support directly the CI service.

    \item Testing infrastructure: this infrastructure aims to provide
        a stable
        environment for users where they can preview the software and services
        developed by INDIGO-DataCloud, prior to its public release.

    \item Preview infrastructure: where the released artifacts are deployed and
        made available for testing and validation by the use-cases.
\end{itemize}

\section{DevOps adoption from user communities}
\label{appendix:devops:users}

\begin{figure}
  \centering
  \includegraphics[width=\textwidth]{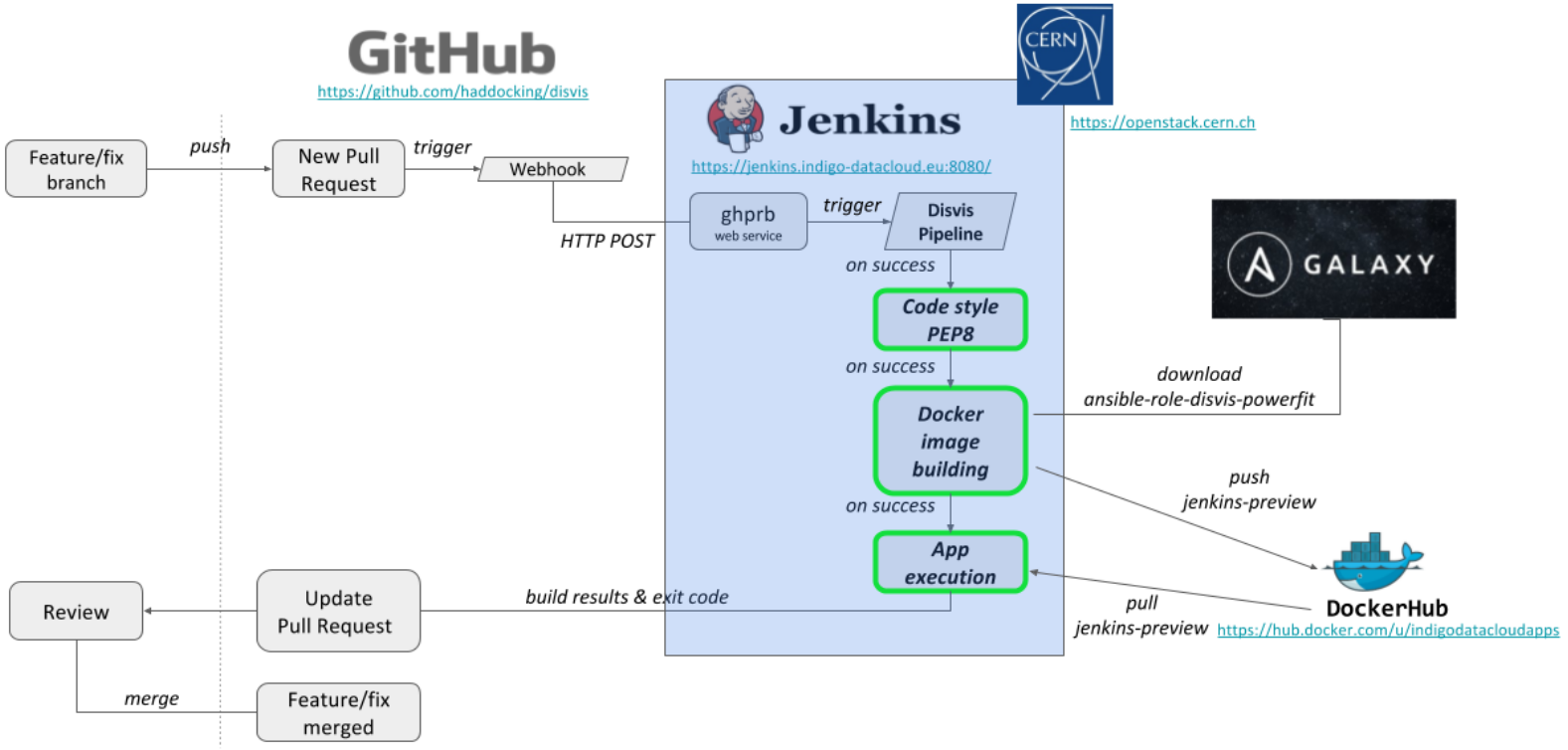}
  \caption{DisVis development workflow using a CI/CD approach}
  \label{fig:16}
\end{figure}

DisVis \cite{disvis} and PowerFit \cite{powerfit} applications were integrated into a CI/CD pipeline described above. As it can be seen in the Figure
\ref{fig:16}, with this pipeline in place the application developers were
provided with both a means to validate the source code before merging and the
creation of a new versioned Docker image, automatically available in the
INDIGO-DataCloud???s catalogue for applications i.e. DockerHub???s {\tt
indigodatacloudapps} repository.

Once the application is deployed as a Docker container, and
subsequently uploaded to {\tt indigodatacloudapps} repository, it is
instantiated in a new container to be validated. The application is then
executed and the results compared with a set of reference outputs. Thus this
pipeline implementation goes a step forward by testing the application
execution for the last available Docker image in the catalogue.

\end{appendices}

\end{document}